\pdfoutput=1
\documentclass[12pt]{elsarticle}%
\usepackage{amsfonts}
\usepackage{amsmath}
\usepackage{amssymb}
\usepackage{graphicx}
\providecommand{\U}[1]{\protect\rule{.1in}{.1in}}
\journal{--}

\begin{document}

\begin{frontmatter}%


%

\title
{The Theories of Relativity and Bergson's Philosophy of Duration and Simultaneity During and After Einstein's 1922 Visit to Paris}%

\author{C. S. Unnikrishnan}%

\address
{Tata Institute of Fundamental Research, Homi Bhabha Road, Mumbai 400005, India\\
E-mail address: unni@tifr.res.in}%

\begin{abstract}%

In 1922, Albert Einstein visited Paris and interacted extensively with an
illustrious section of the French academia. In overfilled sessions at the
Coll\`{e}ge de France and the Sorbonne, Einstein explained his theories of
relativity, and prominent physicists, mathematicians and philosophers
listened, debated, questioned and explored facets of relativity. The 1922
visit had its echoes in the life and works of many who participated,
particularly decisive for Einstein and the philosopher Henri Bergson. This
essay examines that eventful visit, focusing on
the physical and logical aspects of Bergson's critique, with physics commentaries, linking prominent French
physicists and mathematicians Langevin, Painlev\'{e}, Hadamard, Becquerel,
Sagnac, and Kastler. I give particular attention to the logical and empirical
accuracy of the physics issues involved, delineating Bergson's exact reasoning
for his philosophical enthusiasm in Einstein's theory and for the ensuing critique.
Bergson's philosophical stand on duration and simultaneity is
reassessed in the context of later developments in cosmological physics as well as the wealth of empirical data involving comparison of atomic clocks.
Finally we are led naturally to a surprising completion of the philosopher's
program on universal time, duration and simultaneity, in harmony with the
time of the physicist. 
In the appendices after the main text I also give the physics background and
easily verifiable proofs for the assertions made in the text, pertaining to
relativity, simultaneity and time dilation, clearly distinguishing beliefs and
facts.

\end{abstract}%


\end{frontmatter}%



\section{Introduction}

Einstein was invited for an academic visit to Paris in 1922 at the initiative
of Paul Langevin, professor of experimental physics at the prestigious
Coll\`{e}ge de France. Einstein accepted the invitation, after an initial
refusal for reasons of `solidarity to his German colleagues', in an atmosphere
of lingering nationalism after the First World War. He visited alone for a few
days during March-April 1922. In several eagerly and enthusiastically attended
sessions, French academia, from established professors to students, and a few
from the general public got the opportunity to listen to Einstein's exposition
of his theories of relativity. Then they actively sought clarifications and
debated the physical, mathematical and philosophical aspects of the theories.
On the whole, the well-publicised sessions at the Coll\`{e}ge de France and
the Sorbonne were very participatory and by and large friendly.

After the solar eclipse expeditions in 1919 that confirmed the gravitational
bending of light, Einstein was a public figure all over the world. However,
the wounds of the First World War had polarized the world and Europe, most
strongly. Nationalistic attitudes permeated opinions on scientific matters as
well. Einstein was a Swiss-German physicist working in Berlin, for the rest of
the world. Admiration and opposition to Einstein's theories in France were
influenced by these factors. The visit nevertheless happened by the conscious
efforts of some on both sides, who insisted on reconciliation.

Perhaps the only incident that went beyond academic controversy was the
planned boycott of the `German' Einstein's scheduled visit to the Acad\'{e}mie
des Sciences by some members. Forewarned, Einstein cancelled the visit.

There are several descriptions of the 1922 visit and its analysis from
different points of view, but most are centred on one of the sessions held at
the Sorbonne on 6th April, 1922, arranged by the French Philosophical
Society~\cite{Paris-debate,Nordmann,Biezunski,Canales,Latour}. This visit was
particularly engaging for historians because of the encounter between Albert
Einstein and Henri Bergson. In the session at the Sorbonne, Bergson, the
prominent and one of the influential philosophers of modern times and the
professor of philosophy at the Coll\`{e}ge de France, made a few critical and
relevant comments on the notions of time and simultaneity in Einstein's
Special (Restricted) Theory of Relativity. Bergson made these brief comments
at the request of his colleagues, unplanned, despite his deep involvement and
studies of the notions of time, duration and simultaneity~\cite{Paris-debate}.
His detailed opinions and stand on theses notions in the context of relativity
and philosophy are explained with clarity in the monograph `Dur\'{e}e et
Simultan\'{e}it\'{e}: \`{a} propos de la th\'{e}orie d'Einstein' (abbreviated
D-et-S, here), published later, in the same year~\cite{DetS}. At the session,
he mentioned only the need to reconcile the common sense notions of universal
time and simultaneity with Einstein's, in Special Relativity, and said he
believed that this would be possible with a reinterpretation of the theory.
Perhaps he was misunderstood, due to the brevity of the comments. The general
impression was that he raised doubts on the physical relevance of Einstein's
theory, and criticised it as a mere metaphysical thesis with mathematically
defined, but unphysical, multitude of times. Anyway, it is a fact that
Bergson's brief comments reverberated far outside the lecture hall of the
Sorbonne, influencing even the Nobel committee's assessment of the theory; the
1921 prize was given to Einstein in December 1922, primarily for his theory of
the photoelectric effect and there was explicit mention of the
`epistemological nature of the theory of relativity' in the presentation
speech by the Nobel laureate scientist Svante Arrhenius: ``Most discussion
centres on his theory of relativity. This pertains essentially to epistemology
and has therefore been the subject of lively debate in philosophical circles.
It will be no secret that the famous philosopher Bergson in Paris has
challenged this theory, while other philosophers have acclaimed it wholeheartedly.''

In the next several sections, we retrace some instances of the 1922 visit in
the context of the Einstein-Bergson interaction. Our interest here is not the
general historical or socio-political aspects, which are well documented and
analysed~\cite{Paris-debate,Nordmann,Biezunski,Canales}. Our focus is the physical theory, and its consistency and logical
integrity, all in the backdrop of empirical facts. Thus, we necessarily
include the events from the other sessions during the visit, at the
Coll\`{e}ge de France; there were four detailed sessions there that discussed
Einstein's theories (March 31, April 3, 5, and 7, 1922). This then allows the
proper understanding of Bergson's critique and its relation to his philosophy
of time. The 1922 visit serves as an inclusive gathering that brought up and
discussed the core ideas and views on the theories of relativity and the
physics of time. Einstein and Bergson had continued interactions in the
context of a transforming Europe and the world, especially in the context of
the League of Nations. Their relationship was bound to be complex, both
because of their individual views on the political issues involved, and due to
their divergent views on time and simultaneity, which polarized their
acquaintances as well. Due to the opinionated and polarized atmosphere, a
careful and rigorous analysis of the physical and logical issues that were
central to the debate during the Paris visit is lacking. Perhaps it was taken
for granted by most commentators that Bergson's views were flawed because he
failed to understand the relativity theory, new concepts of space and time,
and their mathematical underpinning in the Lorentz transformations. Those who
have read Bergson's work in Dur\'{e}e et Simultan\'{e}it\'{e} (D-et-S)
carefully, especially the sections on the Lorentz transformations and the
discussion of synchronization of clocks and simultaneity, will realize that
the factual situation is the opposite. His rigour as a mathematically
competent philosopher addressing questions in a physical theory was stringent,
and he was meticulous in his analysis. The 1922 encounter between Einstein and
Bergson and the continued debates between Einstein's supporters and Bergson
are documented in the edited volume `Bergson and the Evolution of Physics' by
P. A. Y. Gunter~\cite{Gunter}. However, a careful analysis of the logical
integrity of the positions of physics taken by the adversaries is still lacking.

We will not micro-analyse the actual exchanges during the 1922 sessions,
because they were too brief and incomplete for a proper assessment of the
positions and their reasons. Of course, these exchanges form the basis of the
topics we discuss. We want to focus on the very issues that prompted Bergson
to interfere -- time, duration and simultaneity -- in the context of important
developments in the trajectory of physics. We limit our attention to a few of
those who were present in these sessions, with occasional and necessary
mention to Henri Poincar\'{e}, who was no more, except as a strong ethereal
presence in the intellectual atmosphere discussing relativity and
time.\footnote{Poincar\'{e} passed away in 1912. He was the professor of
mathematics, physics and astronomy at the Sorbonne, member of the French
Academy of Sciences and the Acad\'{e}mie Fran\c{c}aise, apart from serving as
the Chief Engineer at the Corps des Mines. As a member of the Bureau de
Longitudes he studied the problem of the global synchronization of time and
made decisive founding contributions to the principle and theory of
relativity.} The characters, apart from Einstein and Bergson, are Paul
Langevin (host and the master of ceremonies, and `initiator' of the time
dilation problem known later as the `twin paradox', in 1911), Charles Nordmann
(astronomer, physicist, author, and `guide' to Einstein during the 1922
visit), Paul Painlev\'{e} (mathematician and politician who served as French
war minister and prime minister), Jacques Hadamard (mathematician), Jean
Becquerel (physicist and strong defender of Einstein's theory), Georges Sagnac
(inventor of the Sagnac interferometer to prove in 1913 that `ether existed'
and Einstein's theory was flawed), and Alfred Kastler (then a first year
student and later Nobel laureate for his work in optical pumping in atomic
physics). We will leave out many details and other characters, but we note
that the Einstein-Bergson affair echoed for several years through many people
interested in physics, philosophy and politics, and its discussion even after
a century may provoke unexpectedly strong and polarized reactions!

\section{Einstein, Bergson, and Time}

Bergson's studies on time, space, duration and simultaneity in the context of
Einstein's theories were mature when the critical encounter happened. For
Bergson, who had developed a philosophical framework for time and duration,
such an analysis was on his natural philosophical itinerary. He stated about
the origin of the work in the preface to the book,

\begin{quote}
We started it solely for our own benefit. We wanted to find out to what
extent our concept of duration was compatible with Einstein's views on time.
\end{quote}

Then he directly came to the important point that would define the focus,

\begin{quote}
Our concept of duration was really the translation of a
direct and immediate experience. Without involving the hypothesis of a
universal time as a necessary consequence, it harmonized quite naturally with
this belief. It was therefore very nearly this popular idea with which we were
going to confront Einstein's theory. And the way this theory appears to come
into conflict with common opinion then rose to the fore...
\end{quote}

In this section we discuss the physics background to the main points of
contention between Einstein and Bergson. The new notions of time and
simultaneity that evolved from H. A. Lorentz's theory of electrodynamic
relativity formulated during 1895-1905~\cite{Lorentz,Poincare1}, and from
Einstein's own version of relativity in 1905 with differing
interpretation~\cite{Eins-1905}, are at the basis of the debates. Lorentz's
relativity theory, completed and interpreted by Poincar\'{e}, deals with the
modification of the spatial and temporal intervals due to the motion through
the `ether', the hypothetical all-pervading medium for electromagnetic
phenomena and the propagation of light waves. Einstein's theory also is based
on identical mathematical structure, but with radically different
interpretation, without the ether or any universal reference for motion. We
explain the main elements to set the stage. There were two distinct issues
that form the main body of the debate. One was the speed-dependent slowing
down of clocks in motion, or the \emph{time dilation} and resulting
\emph{multitude of times} in relativity theories. The other was the criterion
for the temporal simultaneity of two events that happen at two spatial locations.

\subsection{Two theories of relativity}

We need to discuss briefly the main tenets of the Special Theory of Relativity
(STR) and its consequences to duration and simultaneity before going ahead.
STR originated in the fertile soil of experimental results in optics and
electrodynamics, and the Lorentz-Poincar\'{e} Theory of Relativity (LPTR),
developed during 1895~-1905~\cite{Lorentz,Poincare1}.

The definite completion of the theory of electrodynamics and the
identification of light as the propagating waves of the electromagnetic fields
were the high points in 19th century physics. It was assumed that the
propagation of light required a universal medium, called the \emph{ether}. The
Galilean principle of relativity, that the state of uniform motion cannot be
detected and distinguished from a state of rest, was at the basis of dynamics.
This is related to the undetectability of pure `space'. The question naturally
arose whether the hypothetical ether could be detected by measuring the
relative velocity of light while the laboratory on the earth was moving
through the ether, just as one could detect the presence of the medium of air
by measuring the velocity of sound relative to an observer moving and chasing
the sound waves. However, the enormous velocity of light makes this task very
difficult at multiple levels. The solution employing an interferometer, in
which changes in distances much smaller than a millionth of a meter could be
measured using waves of light, was invented by A. A. Michelson. From a refined
experiment in 1887, Michelson and collaborator E. W. Morley announced the
\emph{failure to detect the motion of the earth through the ether}, in spite of
sufficient sensitivity of the experiment~\cite{M-M}. It was as if \emph{the
ether was as undetectable as the empty space}; the Galilean principle of
relativity seemed to be valid more generally, including all phenomena in
mechanics and electrodynamics.

This null result implied an invariance property of Maxwell's equations of
electrodynamics, similar to the invariance or the equivalence of Newton's laws in
mechanics for all uniformly (inertially) moving observers. Starting with the
ad hoc hypothesis of length contraction (that all extensions contract in the
direction of motion by a certain fraction determined by the velocity through
the ether), H. A. Lorentz eventually arrived at the `Lorentz transformations'
of spatial coordinates and time that achieved this invariance, \emph{while
preserving the invisible ether and the Galilean notion of the variable
relative speed of light}~\cite{Lorentz}. Thus, in Lorentz's theory, the ether
was like a universal absolute reference medium that remained undetectable.
Real motion was motion relative to the ether. A measuring rod (scale) that
moved relative to the ether contracted (length contraction) and the time
measured by a moving clock progressed slower (time dilation). Henri
Poincar\'{e} provided decisive mathematical completion to this theory and also
clarified the meaning of the particular term of `local time' in the Lorentz
transformation, being linked to the synchronization of clocks at different
places in the direction of the motion~\cite{Poincare1,Poincare2}. Poincar\'{e} stated the universal principle of relativity, applicable to all phenomena including gravitation.

That was when Einstein published his ideas of relativity in 1905, with the
same principle of relativity as one of the postulates and a very different
interpretation of the Lorentz transformations~\cite{Eins-1905}. The principle
of relativity is a generalization of Galilean relativity that asserts the
impossibility to distinguish between the state of rest and the state of
uniform motion. Einstein's generalization was to include all physical
phenomena in its scope, just as Poincar\'{e} did. In the words of Max
Planck~\cite{Planck-8},

\begin{quote}
The gist of this principle is: it is in no wise possible to detect the
motion of a body relative to empty space; in fact, there is absolutely no
physical sense in speaking of such a motion. If, therefore, two observers move
with uniform but different velocities, then each of the two with exactly the
same right may assert that with respect to empty space he is at rest, and
there are no physical methods of measurement enabling us to decide in favour
of the one or the other.
\end{quote}

The different interpretation of the Lorentz transformations came from the
second and the characteristic postulate of Einstein's theory -- that the
relative velocity of light is always an invariant constant, relative to any
observer, moving or at rest. Once this non-intuitive feature is assumed,
Lorentz transformations follow as a consequence. Of course, the null result of
the Michelson-Morley experiment is then easily explained and all other
consequences that Lorentz was struggling to understand as physically linked to
the properties of the ether follow as the mathematical consequence of the
theory without the ether. But there was a special price to pay; without any
universal reference, the theory became symmetrical and reciprocal between any
two observers in uniform relative motion. Any observer `A' can claim being in
a state of rest and that the other one `B' is moving, with the Lorentz
modifications affecting the clocks and rulers of only the observer B. In turn,
B can equally well claim that he is at rest and A is moving with exactly the
same modifications happening only to A's clocks and rulers. The fastest clock
and the longest ruler are always in the frame at rest. But, the frame at rest
is equally A and B, from their own frame. This is the point of departure in
Einstein's theory regarding the notions of space and time postulated in
physical theories until then. In other words, either A or B (or any other
inertial frame) can claim the status of the special rest frame, which only the
ether frame could claim earlier. This means that there is no meaning to the
questions, `which frame is really moving?', `which clock and ruler are
actually affected by motion?' etc. In the chosen rest frame, clocks and rulers
are not affected. If this is seriously clashing with the intuitions and common
sense built up over time, that is precisely the uneasiness that troubled many
philosophers and even physicists.

So, in Einstein's theory, there is no physical reason for the modifications of
time and length. In fact, there is no reason, in the sense of a cause-effect
relation, because the cause -- presumably motion -- is not real, but only a
relative notion in Einstein's theory. Relative inertial motion is a totally
symmetrical notion among two reference frames, with equal right to claim the
state of rest. What is demanded is only the consistency within each reference
frame. Since observer A can only imagine what B measures, using theory, and
cannot measure in B's place, mixing measurements does not happen. All they can
do is to compare their physical measurements when they are at the same place -
in infinitesimal proximity.

At the session in the Sorbonne, Bergson started his comments by expressing his
admiration for Einstein's theory, which was `new physics and in some respects,
a new way of thinking'. He briefly stated his opinion that there was nothing
incompatible between the common sense notion of a single universal time and
Einstein's relativity~\cite{Paris-debate};

\begin{quote}
Common sense believes in a unique time, the same for all beings and all
things...the idea of a universal time, common to conscious beings and to
things, is a simple hypothesis.

But it is a hypothesis that I believe to be founded and which, in my opinion,
contains nothing incompatible with the theory of relativity. I cannot
undertake to demonstrate this link. It would be necessary to study real
duration and measurable time much more minutely than I have just done. It
would be necessary to take the terms which enter the Lorentz' equations one by
one and search for their concrete significance. Then one would find that the
multiple times the theory of relativity deals with are far from all being able
to pretend the same degree of reality...

But all that I cannot establish as regards time in general, I seek your
permission to do, or at least glimpse, for the specific case of simultaneity.
Here we will see easily that the relativistic point of view does not exclude
the intuitive point of view, and even necessarily implies it.
\end{quote}

Thus, only the issue of simultaneity was elaborated by Bergson (what he called
a `glimpse') in the discussion at the Sorbonne, due to paucity of time. We
will thoroughly analyse the details in the section `Simultaneity: Einstein vs Bergson'.

Even a cursory reading of D-et-S will convince anybody about the rigorous and
clear nature of Bergson's writings. Penetrating the simple mathematical
aspects of the relativity theories of Lorentz, Poincar\'{e} and Einstein was
elementary for Bergson, because of his background in mathematics and proven
ability and record, before he chose philosophy as his subject at the \'{E}cole
Normale Sup\'{e}rieure. We have to keep in mind the important fact that in
1922, there was no direct experimental demonstration of time dilation as an
observable physical effect. Hence, \emph{the dispute was not about whether
effects like time dilation could happen in nature or not; the debate was
strictly about whether the equivalent and symmetrical role to the two
observers in relative motion characteristic of the Special Theory of
Relativity was consistent with an actual asymmetrical physical effects
deduced}, favouring one of the clocks as slower than the other, without a
justified reason. After all, it was known to Bergson (and discussed in detail
in the first chapter of D-et-S) that the time dilation of clocks in motion is
central to Lorentz's non-symmetrical theory of relativity in the ether,
formulated well before Einstein's theory. If this crucial point is missed,
then Bergson's stand is easily misunderstood and misrepresented.

\subsection{A mysterious postulate}

The propagation of light was central to the development of the theories of
relativity. A feature that was of some uneasiness to many, if not most, was
the fundamental characteristic postulate of Special Relativity, that the
relative velocity of light was an invariant constant for all uniformly moving
(inertial) observers. The discussions on simultaneity are intertwined with
this topic. This may be best described, in the context of the visit, by
quoting the uneasiness even in an admirer, friend, and supporter, Charles
Nordmann~\cite{Nordmann},

\begin{quote}
However, ... there is still something infinitely troubling in the
Einsteinian system. This system is admirably coherent, but it rests on a
particular conception of the propagation of light. How are we to imagine that
the propagation of a ray of light could be identical for an observer who flies
away from it, and for an observer who rushes forward to meet it? If this is
possible, it is in any case inconceivable to our customary mentality, and no
matter how hard we try, we cannot make the mechanism and nature of that
propagation intelligible.

It must be confessed that here lies a mystery which eludes us. The whole
Einsteinian synthesis, as coherent as it is, rests on a mystery, exactly like
the revealed religions.
\end{quote}

The invariance of the relative speed of light, an `infinitely troubling'
mystery indeed, is the core postulate of Special Relativity. It is obvious
from Nordmann's statement quoted above that the invariance of the velocity of
light was not an empirically proven fact, contrary to the general erroneous
(modern) belief that the `null' result of the Michelson-Morley experiment of
1887 was the `proof' of this 1905 postulate\footnote{The Michelson-Morley experiment was an optical proof for the principle of relativity. Its null result is obviously consistent with the Lorentz-Poincar\'{e} theory with the ether and Galilean light as well as with Einstein's theory. After all  Lorentz explained the null result in 1894 using length contraction in ether.} As in 1922, even today we do not
yet have any experimental confirmation of this `postulate of revelation' (see
later for details). However, this was not an issue of discussion or contention
during the 1922 visit. Bergson understood clearly that the velocity of light
was not a true invariant in Lorentz's ether relativity (`half-relativity', in
his book), whereas it appeared to be a universal invariant in Einstein's
theory, and accepted as a possibility, in spite of its anti-common sense
nature. The difference in the nature of propagation of light in the two
theories is at the very basis of their different notions of simultaneity of
spatially separated events. I stress again that it is an empirically
unverified postulate, as evident in Nordmann's clear statement, contrary to
the belief held by most physicists that it was verified in experiments even
prior to Einstein's theory. Though it was not discussed during the 1922
sessions, it figured prominently in the published debate later, between
Einstein's supporters and Bergson~\cite{Gunter}. I quote from an article by
Andr\'{e} Metz (a `faithful disciple' of Becquerel's\footnote{Becquerel's role
in the continued debate with Bergson on the twin paradox in Einstein's
relativity is discussed later.}) in Revue de Philosophie (1924), ``Einstein
arrived at a different result by starting from an experimental fact...the
isotropy of the propagation of light...the isotropy is firmly established, not
only by the Michelson's experiments, but also, and especially, by all the
experimental verifications of electromagnetic theory (Maxwell's theory)''. 
Only people who have not read and understood the work of Fresnel, Fizeau, and
Michelson, and the Lorentz-Poincar\'{e} relativity of 1895-1905, could make
such blatantly fallacious statements. Consistency cannot be equated to
empirical proof in science.

A. A. Michelson published a resume of his researches as a monograph `Studies
in Optics' in 1927~\cite{Michelson1927}. One chapter is dedicated to the
measurements of the velocity of light, in which the Michelson-Morley
experiment is not even mentioned! In that chapter, he clearly states that the
constancy of the relative speed of light is a \emph{postulate} of the theory
of relativity. And in the chapter on `Relativity', he discusses the
Michelson-Morley experiment, correctly, as a test of the postulate of the
general principle of relativity, which asserts the impossibility to detect the
earth's motion through the ether. He concluded the chapter by saying that the
result of his later experiment (with A. H. Gale) `could be considered as
additional evidence for relativity, or equally as evidence of a stationary ether'.

The hypothesis of the constancy of the speed of propagation of light relative
to any inertial frame has remained untested experimentally to this day
precisely because it is intertwined with the issue of synchronization and
simultaneity of clocks in two different locations; the measurement of the
velocity necessarily requires the measurement of duration that light takes to
propagate from `here' to `there', and hence, between two clocks. Therefore,
all measurements, direct and indirect, have been limited to the two-way speed
in which light is brought back to the same point in space, to measure the
duration of the two-way propagation, as in the Michelson-Morley experiment.%

\begin{figure}[ptb]%
\centering
\includegraphics[
width=5.0in
]%
{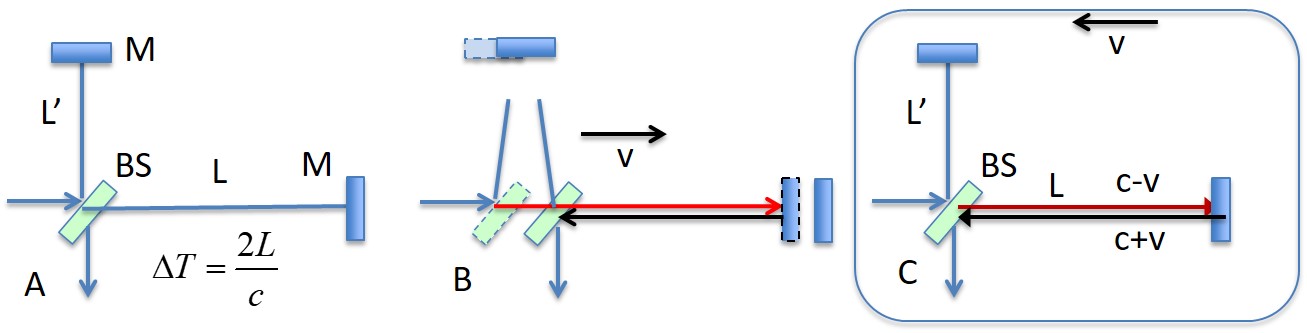}%
\caption{A: The Michelson Interferometer. Light waves travel from the beam
splitter BS to the two mirrors M and back and exit overlapped, causing
interference and `fringes'. Any differential change in the distances to the
mirrors results in a change of the intensity of the light output, or a shift
of interference fringes. B: In the Michelson-Morley experiment, the
interferometer is in motion with the earth's velocity $v$ while the light
waves are on their two-way trip. As seen from a global frame, light
traverses unequal up and down distances in the direction of the motion, as
indicated with the red and black arrows. C: In the `comoving frame' or the `rest frame' of the earth and the interferometer, the relative velocity of light in the two directions is $c-v$ and $c+v$. }%
\label{MM}%
\end{figure}

It is \emph{very easy to see why the Michelson-Morley experiment cannot decide on
the isotropy of the propagation of light}; the time taken in propagation over a
length $L$ at speed $s$ is $L/s$. So, if the speed is not isotropic, but
instead $c-v$ while light is in one direction and $c+v$ in the return path
after reflection in the two-way experiment, total duration taken is
\[
T=\frac{L}{c-v}+\frac{L}{c+v}=\frac{2L}{c(1-v^{2}/c^{2})}%
\]
Hence, the total duration is different from the isotropic value $2L/c$ only by
the small `second order' term, $(1-v^{2}/c^{2})$. If both length and time
(clocks) are modified due to the motion (through ether, say) by factors
$\sqrt{1-v^{2}/c^{2}}$, this excess factor of $(1-v^{2}/c^{2})$ is exactly
cancelled and we get $2L/c$ as the duration even though the propagation of light
is not isotropic! On the other hand if one postulates that the speed of light
is isotropic in any frame, then there cannot be any modification of length or
time in such frames. Either view is consistent with the Michelson-Morley
experiment, making such experiments indecisive on this issue. It is an
elementary fact, and there is no place for the erroneous belief.

It is obvious that a measurement of the one-way relative speed of light needs
to tackle the hard problem of synchronizing two separated clocks. However,
even the seasoned experts do not always realize the crucial point and one can
see a number of well-cited publications claiming the verification of the
postulate in measurements involving one-way propagation of light, with clocks at two
different locations. The irony and embarrassment of the situation will be
understood if it is noted that the need for such one-way measurements, to
demarcate Einstein's theory from the Lorentz-Poincar\'{e} theory, was
reiterated in a paper in the journal Physical Review Letters by M. Ruderfer in
1960~\cite{Ruderfer1}. He suggested an experiment based on the newly
discovered M\"{o}ssbauer effect. Very soon Ruderfer published a vital
erratum~\cite{Ruderfer2}, when he understood the fundamental issue of the
inseparability of the propagation delay and the duration measured with
separated clocks. This fundamental `catch' was already known and discussed by
Poincar\'{e}, stressing the role of `convention' in synchronization, but it is
a subtle and deep issue that is easily missed, as history shows. Oblivious to
the content of the erratum, researchers went ahead \cite{Champ-blunder} and `verified the
fundamental postulate' and `ruled out the Lorentz-Poincar\'{e} theory', while
citing both Ruderfer's proposal paper as well as the nullifying erratum! And
many are continuing in vain to refine these tests on light \cite{Krisher-Will}, still remaining in
the dark about the interdependence of synchronization and time dilation of
separated clocks in moving frames, and the inability of such tests to
decide between the two kinds of theories, as pointed out clearly by Ruderfer.

\subsection{Bergson and Einstein's theory}

Simultaneity of physical events as a fundamental premise was discussed in
detail by Einstein, with the scenario of an observer on a platform and another
in a moving train, in his expository monograph on the theories of relativity
(1916), `Relativity: The Special and General Theory'~\cite{Eins-book}. Bergson
meticulously analysed Einstein's conception of simultaneity, in D-et-S.
Contrasting the common notion of simultaneity with that in the theory of
relativity, the philosopher hoped that there would be convergence of the
concepts. But, it seemed that there was irreconcilable discord between the two
views. Bergson pointed out the inconsistency between the conclusions arrived
at by Einstein and the symmetric reciprocity of Special Relativity (this
easily verifiable point has been discussed by a few others ever since, but
generally ignored. We will discuss our transparent proof later, in the section
on simultaneity). He also discussed in great detail the problem of the two
brothers and their clocks (the twin paradox), and the issue of multitude of
times in relativity. The relativists defended strongly and vigorously (`more
Einsteinian than Einstein'), concluding that Bergson misunderstood the theory
and the Lorentz transformations. The relativists went further and concluded
that the edifice of the Bergsonian philosophy of time and duration crumbled
due to his mistaken views about real time, which was the time of physics as
well as of Einstein's theory.

There is no need to examine in detail the philosopher's distinction between
the notions of `time' and `duration' in our discussion. But it is important to
make a clarifying remark. Only the `duration of time' is physically (and
psychologically) sensible. Duration is the measure of elapsed time, and all
physical phenomena are events measuring some duration. Thus, the concepts of
continuity and extension are naturally built into the notion of duration. As
Bergson wrote in D-et-S, his notion of duration harmonized naturally with the
hypothesis of a universal time. Bergson's examination of the physical theory
was to see whether it was in harmony with the time in physics.

Bergson's position on Einstein's theory and the motivation for his hopeful
admiration for the theory is clearly stated in D-et-S, without any scope for
ambiguity, right after the first chapter `Half Relativity' and two pages into
the second, `Complete Relativity'. In `Half Relativity' he analyzed and
interpreted the terms in the Lorentz transformations, one by one (as he
indicated in the session at the Sorbonne), and discussed in detail the
synchronization of clocks and its dependence on the protocol with the
propagation of signals. This was in the context of the Lorentz's theory, and
the interpretation was close to that of Poincar\'{e}, though Bergson did not
(surprisingly, for me) mention Poincar\'{e} in D-et-S. With Lorentz's theory
and its ether as the privileged frame, there was one system (of reference) S
that possessed the master clock and time, and in a system S' that was
relatively moving there was a different, dilated real time. However, there was
no physical way to know whether one was moving or not through the insensible
ether's privileged frame! \emph{Hence, Lorentz's relativity was at direct
conflict with a single universal time central to Bergson's philosophy.} With
Einstein's relativity, all observers S, S', S\textquotedblright..., were
deemed equivalent with equal right to claim the state of rest; there was
multitude of times, now apparent and interchangeable, with the dilation always
affecting the `other' clocks. Therefore, \emph{Bergson hoped and reasoned that
these mathematical multitude of times could not be identical to the real time,
and the whole picture could be made consistent with the Bergsonian preference
for a single universal real time}. This is the crux of Bergson's evaluation
and interpretation of Einstein's relativity.

As Bergson stated in the preface to D-et-S, ``we started it solely for our own
benefit''. We see that Bergson approached the theory as a charmed suitor and
not as a critique, but ended up criticizing the theory to bring out what he
thought was its interpretational essence that would harmonize with a universal
real time. \emph{Lorentz's theory, with its real time dilation for real motion
in invisible ether, was in direct conflict with Bergson's concept of time,
whereas Einstein's theory without an absolute reference for motion gave him
the hope of reconciling the notions of time in physical relativity and time of
the philosopher, and indeed of common intuition.} With no empirical method
known to choose one theory over the other, the philosopher had a choice in
Einstein's theory, with the possibility of reconciliation with proper
interpretation. If Lorentz-Poincar\'{e} theory were the only theory, Bergson's
thesis on universal duration and simultaneity would already be invalidated,
being in conflict with the only physical theory of empirical experience. This
is the core point concerning Bergson's interest and engagement, as a
philosopher, in the physical theories of relativistic time.

\section{Irrelevant Common Sense}

When we discuss the Special Theory of Relativity, we will need to disregard
certain kinds of objections that make sense from the point of view of common
sense, but that are not immediately useful in the analysis of the theory. Two
instances can give a flavour of the situation.

I am an observer at rest in STR - I am always at rest in my frame in STR, by
definition; that is my rest frame. I see several things moving past.
Surprisingly, they all are moving in the same direction and with the same
velocity. An analogous thing happened in history when it was noticed that the
entire system of stars and the globes of the solar system alike were moving as
an ensemble from east to west every day. This was finally simplified by
realizing that the earth is `really' rotating, while stars stayed more or less
fixed. So, should I conclude from common sense that I am the one who is really
moving and physical effects should happen only to my clocks and measuring
rods, if at all? STR insists that one is not justified to draw such a
conclusion -- there is nothing called real uniform motion. Every inertial
motion is equivalent to a state of rest. In one's rest frame, however humble,
the rest of the universe is moving; the rest of the clocks and rulers change.

More down to earth, I see a flock of birds in the standard formation, but all
going backwards. I know that it is impossible. But, the physical theory of
space and time doesn't care. As physical objects, the birds are in motion,
while I am at \emph{rest}. To the question, `who's clock is going slower?',
the theory gives the answer: the internal clocks of the birds are running
slower, whether they are flying forward or backward. If they appear still,
despite the flutter, they are also at rest and there is no time dilation. In
the ether-relativity, it would have been different; the appearance of the
`unnatural' motion of the birds must be due to my faster real motion relative
to the ether. Can I say that the birds that are `really' moving forward slower
than me, and hence appearing moving backwards, must also be ageing faster
relative to me, guided by common sense? Einstein's theory answers in the
negative, because it is based on kinematics in the empty space. Since
durations cannot be compared unless I mark the time on the clocks twice -- at
the start and at the end, at the same place, there seems no way to verify my
common sense guess with birds that have flown past. (However, see appendix II).

So, the distinction is clear. In LPTR, modifications of lengths and durations
are real physical effects that increase with the velocity relative to the
ether. In STR, these modifications depend on the relative motion between the
frames and the choice of being at rest is left free; only clocks and rulers
external to our rest frame are affected. We are at rest, in our frame, and
nothing happens to our clocks and rulers. Anybody's clock can be chosen to run
the fastest by merely occupying his frame, and claiming the state of rest.
There are as many fastest (or `normal') clocks as there are observers --
hence, the reciprocal multitude of times of the Special Relativity. What is
not noticed is a glaring logical gap; if one could claim a state of rest in
situations of relative inertial motion, then one can also claim a state of
motion at will, making the physics of clocks arbitrary. \emph{Given a value
for the relative velocity between two frames, there is an infinity of possible
partitions of individual velocities between the frames}. But, that is the
topic of discussion in a separate paper on the logical structure of the
theories of relativity~\cite{LogicofRel}.

Clearly, the philosopher might not have been worried about the conflict with
common sense in the way of the examples we mentioned. He was worried about the
meaning of the assertions of the theory and whether they rigourously and
logically follow from the symmetrical structure of the theory, with its
temporal multitude. Einstein, on the other hand, was concerned only about the
consistency of the theory with what was measurable. He was also aware that his
theory gave the same predictions as the Lorentz-Poincar\'{e} ether-based
theory in situations of common physical measurements. Above all, the confirmed
predictions of the General Theory of Relativity (GTR) for the gravitational
physics of the motion of the planets and the bending of light near the sun
were an indication of the correctness of his approach, and was perhaps the
basis of his confident modesty.

\section{Langevin's Twin Paradox}

The Special Theory of Relativity was projected as a theory of space and time,
rather than merely as a theory of motional modifications of distances,
durations, and related physical quantities, after H. Minkowski's mathematical
formulation in which `time' was depicted as a fourth dimension, along with the
three spatial dimensions~\cite{Minkowski}. However, the fact that this
dimension had the imaginary nature, in the mathematical sense, confined the
geometrical theory within the circle of physicists and mathematicians (in
older texts, it was depicted as $(ic)t$, where $i=\sqrt{-1}$). It was through
a lecture by Langevin at the International Congress of Philosophy in Bologna
in 1911~\cite{Langevin-twin} that many philosophers were jolted into a
discussion about the multitude of times in the physical world, one each and
one's own for every inertial observer. Bergson was a speaker at the congress
in Bologna, where he spoke about `Philosophical Intuition'. There, Langevin
mentioned about time dilation of transported clocks in Einstein's theory, and
discussed two concrete examples. One was about the comparison of the lifetimes
of two samples of radioactive Radium in relative motion, and another was the
scenario of the space traveller. He said,

\begin{quote}
Imagine a laboratory attached to the Earth, which motion can be considered
as uniform translation, and in this laboratory there are two perfectly
identical samples of radium. What we know about the spontaneous evolution of
radioactive materials allows us to say that if these samples are kept in the
laboratory, they will lose both their activity the same way over time and
their activities remain continuously equal. But then send one of these samples
with a sufficiently high velocity and then bring it back to the laboratory;
this requires that at least at certain times this sample has undergone
accelerations. We can say that on return, its proper time between departure
and return is less than the measured time interval between these events by
observers attached to the laboratory, so that it has less evolved than the
other sample and therefore it will be more active than the latter; it will
have aged less, having been more agitated.
\end{quote}

As the next example of `lived durations' of the traveller he said,

\begin{quote}
For this it is sufficient that our traveller consents to be locked in a
projectile that would be launched from Earth with a velocity sufficiently
close to that of light but lower, which is physically possible, while
arranging an encounter with, for example, a star that happens after one year
of the traveller's life, and which sends him back to Earth with the same
velocity. Returned to Earth he has aged two years, then he leaves his ark and
finds our world two hundred years older, if his velocity remained in the range
of only one twenty-thousandth less than the velocity of light. The most
established experimental facts of physics allow us to assert that this would
actually be so.
\end{quote}

This second example evolved into the fable of the `twin paradox' of
relativity. One of the twin brothers go out in a long and fast space-voyage.
When he returns home, he has aged much less than his brother who stayed back.
What is the explanation for the unequal ageing? Langevin was aware and careful
about the fact that, in Special Relativity, any observer can claim the state
of rest and the special relativistic ageing was symmetric, unless some
kinematical asymmetry was invoked. He cited the asymmetry of the accelerations
of one of them, required to make the return trip for the eventual comparison
at the same place. He stated that the one who had aged the least is the one
who had suffered the greatest accelerations. He devised a scheme of light
signals for the continuous communication between the traveller and the earth
station and demonstrated the asymmetry as equivalent to the asymmetry of
Doppler shifts of the light signals.

However, there was nothing in the inertial physical basis or the mathematical
structure of Special Relativity that could justify the assertion, `one who had
aged the least is the one who had suffered the greatest accelerations'. Those
who do not agree with Langevin's unjustified reasoning would naturally
interpret this situation as a paradox -- hence the twin paradox, where one
person remains on the earth (A) and his twin brother (B) is the space
traveller. One would argue that in the space traveller's rest frame, it is the
earth dweller who is relatively moving and Einstein's theory (through the
Lorentz transformation on duration) predicts that the earth clocks (A) run
slower, irrespective of the direction of travel and without regard to whether
they are continuously monitored by B. There is nothing in Special Relativity
which demands a special treatment while turning back - that is a small
duration compared to years of travel. So, relative to B, it is A who ages
slower, whereas relative to A, it is B who ages slower. Since both have equal
special relativistic right to claim the state of rest, both are right and the
only logical conclusion, without invoking any new theory, is that neither will
be slower or faster. Hence, time dilation in Special Relativity must be an
unphysical illusion -- that is the essence of the twin paradox and the
critique of Special Relativity. This problem doesn't arise in
Lorentz-Poincar\'{e} theory, where time dilations occur only while moving
relative to the ether -- the situation is asymmetric from the start to end,
without accelerations. In any case, the \emph{Special Theory of Relativity
does not specify a reason for a clock that underwent a brief period of
noninertial motion to go slower than the one that didn't}.%

\begin{figure}[ptb]%
\centering
\includegraphics[
width=2.4302in
]%
{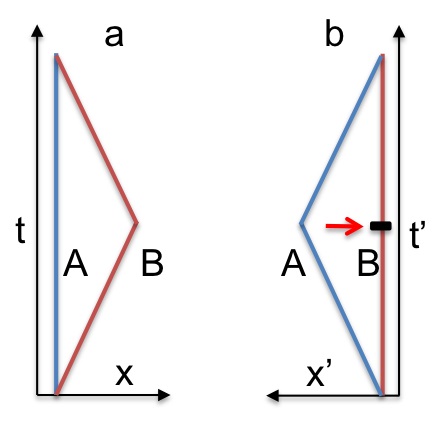}
\caption{Space-Time diagrams of the twin clock problem. a) The diagram from
the rest frame of A, with vertical `world line' in blue. B moves out in space and
returns. The `length' of B's world line and the duration measured in his clock
are shorter, due to the negative contribution from the motion in space. b)
Relativity from B's rest frame. Now B has the vertical world line and A moves
out in space and returns. B feels a squeeze around the midpoint,
indistinguishable from gravity, for a short duration (red arrow). A's world
line is shorter and the proper duration in his clock, is less. The situation
is entirely symmetrical, save for the brief period of gravity in B's world
frame. There are many who draw only the diagram (a) and conclude that
`obviously B ages less', without any mention of the diagram (b), acceleration,
or gravity.}%
\label{Twin-ST}%
\end{figure}

To be precise, the physical time (duration) recorded by a moving clock in
Einstein's theory is given by a simple formula,
\begin{equation}
(d\tau)^{2}=(dt)^{2}-(dx/c)^{2}%
\end{equation}
where $dt$ is the conventional duration of the familiar Galilean time and $dx$
the distance moved. This equation defines the `geometry' of STR. The
difference from the familiar Pythagoras triangle formula is the negative sign.
So, the `time base' of the space-time triangle, corresponding to the duration
in the rest frame, is \emph{longer} than the sum of the other two sides (see
Fig. 2). In the frame of the clock, it is at rest and $dx=0$; then the
physical time and the clock time are identical, and flow fastest. For any
moving clock, the physical time $d\tau$ is shorter than $dt$ because of the
subtractive contribution from moving the distance $dx$. While the rest frame
clock C1 marks 100 seconds, another clock C2 that has relatively moved by a
1000 kilometres marks about 55 nanoseconds less than 100 s, \emph{as
theoretically calculated from the C1 rest frame}. In contrast, C2 is at rest
in its frame, and while it marks 100 s, C2 has moved 1000 km and records 55
nanoseconds short of 100 s, \emph{as calculated from C2 frame}. Since any of
the inertial observers has the right to claim the state of rest in the theory,
$dx=0$, after rejecting a deciding privileged frame like the ether, the clocks
in that frame run the fastest, and all other clocks slower. The arbitrary
choice of the `immobile' frame with complete equal rights is the source of the
controversy. Lorentz and Poincar\'{e} didn't have to face this difficulty
because there was the special frame of the ether in which the immobile clocks
had the fastest rate, and any clock moving in the ether was slower; there was
no reciprocity.

\subsection{Divergent views: \newline Langevin, Einstein, Painlev\'{e}, and
Becquerel}

Langevin's 1911 solution remains as one of the popular answers to the paradox
in modern text books.\footnote{Different text books on Special Relativity
offer three or four physically different solutions to the twin paradox,
signalling the perennial confusion and the lack of consensus. These different reasons 
are Doppler effect, Change of simultaneity, Acceleration asymmmetry, and Gravitational 
time dilation in accelerated frames.} Yet, most of
the discussions do not spell out how exactly the asymmetry of acceleration
translates to an observable physical effect. Instead, nearly everybody insists
that the paradox is cleared entirely within the Special Theory of Relativity.
Few realize that Einstein had rejected such a resolution as incorrect in 1918,
four years before the Paris visit and long after the assertion on asymmetric
time dilation was made in the 1905 paper on the Special Theory of
Relativity~\cite{Eins-1905}. He admitted in the paper published in Die
Naturwissenschaften~\cite{Eins-twin} that the relativists were `evading the
issue' and \emph{the solution required going beyond the special theory,
because the situation was symmetric in the theory}! Most textbook authors who
have written on time dilation in relativity are not even aware of this paper.
Einstein's much awaited and delayed resolution of the paradox was that the
pseudo-gravitational field experienced during the acceleration phase (during
the reversal of the velocity for the return journey) resulted in sufficient
amount of gravitational modification of relative time in the spaceship frame
according to his new theory of gravity -- the General Theory of Relativity --
that there was really no contradiction or paradox~\cite{Eins-twin}. This was
indeed his first definitive and quantitative response to the twin paradox.
Both Nordmann and Kastler mention the need for General Relativity input to
solve the paradox, in their memoirs of the 1922 visit. In contrast, most text
books and monographs do not even mention Einstein's asserted solution and
pretend that the correct solution is entirely within Special Relativity. We
sketch Einstein's simple gravitational solution to the twin paradox in the
appendix I.

The paradox was vigorously discussed during the 1922 visit, brought to debate
by the insistent Painlev\'{e} in one session at the Coll\`{e}ge de
France~\cite{Nordmann,Kastler}. The summary from Nordmann's record is that
Einstein's replies to the queries were satisfactory to the gathering and
Painlev\'{e}. But, Einstein seemed to have chosen to keep quiet on the
necessity of invoking General Relativity and the gravitational time dilation
to justify the asymmetric time dilation, and to resolve the paradox that arose
in the special theory. We will see that while this did not affect Einstein's
reputation, his physicist defenders like Jean Becquerel continued to use
biased and incorrect arguments entirely within Special Relativity to justify
the asymmetric dilation. The irony is exemplified by a letter that Becquerel
sent to Bergson, in which Becquerel `explained away' the asymmetric time
dilation in Special Relativity without discussing acceleration, even during
the reversal of the journey. Little did Becquerel realize that his discussion
of time dilation would become a record of how even distinguished physicists
misunderstood the Special Theory of Relativity and its criticisms, as Bergson
spotted and exposed in an appendix to D-et-S, second edition.

Becquerel's letter was reproduced (to represent his views faithfully) in the
later editions of Duration and Simultaneity, in the first
appendix~\cite{DetS-App}. For Bergson, it was easy to spot and show that
Becquerel was inadvertently using a theory of relativity with a preferred
frame and not Special Relativity in his `demonstration' (Bergson did not name
the `most distinguished physicist' in his book).%

\begin{figure}[ptb]%
\centering
\includegraphics[
width=3.7488in
]%
{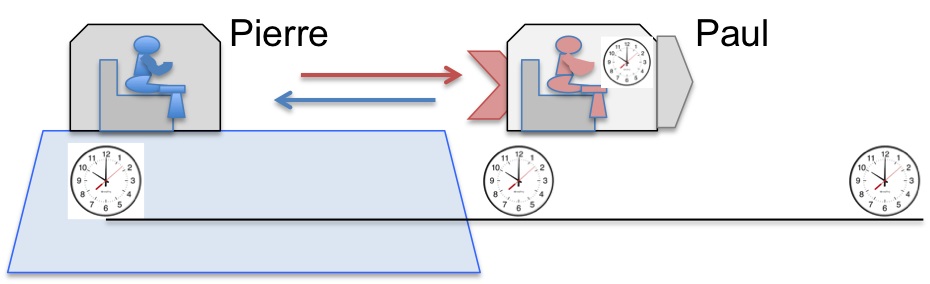}%
\caption{Becquerel's version of the scenario of the twins, where the
asymmetrical ageing is deduced even for the inertial phases, is indicated.
Clocks are synchronized to Pierre's clock. Becquerel asserted that Paul would
see the clocks that move past him read more time relative to his clock, at
rest with him, and hence Pierre would age more.}%
\label{Pet-Paul}%
\end{figure}

Briefly, Becquerel discussed the standard twin problem, with one of them on
earth and the other blasting off at high speed. In those French discussions,
they are named Pierre and Paul. In Becquerel's scheme without accelerations,
there are clocks placed throughout the trajectory, all synchronized with the
earth clock. Paul adjusts his clock to read the same time as Pierre's as the
two meet in their paths. What is then simpler than Paul's just comparing his
clock with the nearest `earth synchronized clock (ESC)' on his path to find
out who ages slower? Becquerel concludes from the Lorentz transformations and
Special Relativity that Paul's clock always reads \emph{less} than ESC. So,
according to Becquerel, Paul in his rest frame would conclude that he was
ageing slower after he separates from Pierre, even without any asymmetry
introduced. Since the total time should be twice that for half the journey,
once it is argued that Paul remained younger than Pierre during half the
journey, Becquerel concluded that Paul would return younger. Becquerel
analyzed the entire journey from the frame of Paul in the spacecraft, i.e.,
Paul's rest frame. And, surprisingly for a physicist, he concluded that Paul
would see less duration elapsed in his clock, compared to Pierre's, even
during the inertial journey. So, Becquerel, in effect, assumes that Paul is
the `really moving guy', relative to Pierre as well as in Paul's own
reckoning. He forgets the most important aspect of Special Relativity that the
clock in its own rest frame goes the fastest and the (proper) duration
recorded in any relatively moving clock is less. Therefore, irrespective of
what the other passing clocks read, Pierre must have clocked less time, having
moved some relative distance. Remember the formula for the elemental
relativistic durations, $(d\tau)^{2}=(dt)^{2}-(dx/c)^{2}$. It was an
elementary mistake from Becquerel. I think his confusion was because he
mistook the clock's local time readings at the spatial locations of
comparisons (as given by Lorentz transformation formula) as identical to
Pierre's elapsed duration.

In contrast, Bergson's analysis is clear; concerning the outward and inward
journeys he says,

\begin{quote}
But, from the standpoint of theory of relativity, there is no longer any
absolute motion or absolute immobility. The first of the two phases just
mentioned then becomes an increasing distance apart between Pierre and Paul;
and the second a decreasing one...The symmetry is perfect...their situations
are interchangeable.
\end{quote}

Then Bergson shows that, using the same Lorentz transformations, it is
Pierre's clock that runs slower, as observed from the rest frame of Paul's,
during the inertial portions of the journey. This was exactly the same
conclusion Einstein emphasized in his 1918 twin paradox paper, for the
inertial portion of the journey. In summary, Becquerel made the grave mistake
to argue that the clocks in the rest frame S were slower than the clocks in a
relatively moving system S' in Special Relativity. This error was carried by
some other French supporters of Einstein's, like Andr\'{e} Metz, in their
criticisms of Bergson, but we will not go into those details.\footnote{P. A.
Y. Gunter's edited volume contains translations of articles by Metz and
Bergson's replies, in Revue de Philosophie (1924).} Becquerel's error is there
in the reprinted letter in D-et-S for all to see and easily verify.
Ironically, those who claim that the twin paradox is resolved entirely within
Special Relativity are inadvertently asserting that Einstein's solution in 1918 \cite{Eins-twin} and
the Equivalence Principle (which he used in the solution) are incorrect,
without citing any reason.

\subsection{The need for logical rigour and consistency}

Discussions on the twin paradox often slip philosophically and logically in
distinguishing time dilation in nature and time dilation in a particular
theory. Whether a theory represents the phenomenon (of asymmetric time
dilation here) in its logico-mathematical structure depends on the nature of
the theory, irrespective of whether there is such a phenomenon in nature.
Bergson, being a philosopher, was arguing that \emph{the symmetric and
reciprocal theory of Special Relativity could not have asymmetrical multitude
of real times}. \emph{The same point was made by Einstein himself in his 1918
paper on the twin paradox}. At a time when direct experimental evidence for
the time dilation was not available, further questions on the reality and the
physical reason for the phenomenon were empirically undecidable. Note that
nobody really debated about the gravitational time dilation as a paradox in
days when it was not verified, though the effect was known for as long as the
twin paradox of STR itself; there was no reciprocity about two different
heights in a gravitational field. But, the problem special relativists faced
was that if they admitted that the time dilation was symmetric in STR, the
theory would become vacuous.

Finally when direct experimental evidence for time dilation became available
in 1941, first in the dilation of the life time of cosmic ray
muons~\cite{Bhabha,Muon-Rossi}, it was clear that acceleration had no direct
role to play in time dilation; the energy loss of these charged particles
subject them to enormous deceleration (more than a trillion times the gravity
on earth), but the time dilation measured is simply related to the velocity
only. However, acceleration that happens only after half the journey is
invoked as the agent of asymmetrical time dilation in all modern discussions
of the twin paradox. Finally, even a layman may be able to ask a `most
distinguished physicist' how would acceleration play any role at all if one
decides to freeze the clock during the brief durations of acceleration and
restart it where it was stopped after the acceleration! After all, the distant
clock with Pierre cannot get affected by the accelerations of
Paul~\cite{Unni-twin}.

Bergson's view of the twin clock problem within Special Relativity was as we
stated already; in the completely symmetric theory, there was no room for the
alleged asymmetry. The analogy he used was the appearance of one of the twins
to the other when they are nearby and when they are far. At a distance, the
person will appear as a midget, but who of the two is appearing as the midget
will depend on who's view is taken as privileged. Paul will appear as midget
from Pierre's point of view and Pierre will be judged as midget from Paul's
point of view. At the end, when both can be together the difference in size
dissolves away. The apparent plurality of durations is similar and in
Bergson's view, there is no real relative time dilation in Special Relativity.
The corollary is that if time dilation was known to be a fact of nature,
Bergson would have probably argued that the explanation was not in Special
Relativity, but in a different theory (like Lorentz's theory) with the
appropriate inherent asymmetry in inertial motion.

We conclude this section on time dilation with one ironical empirical outcome
and a related incident of quenched criticism. An experiment on time dilation
`on a set of twins' was done with Caesium atomic clocks in
1972~\cite{Hafele-Keating}. (In this experiment, there were also gravitational
effects on the rate of clocks, but the value was known from General Relativity
and could be subtracted to get the ageing due to only the relative motion).
One set of the twin clocks (Pierre) remained in the Naval Observatory in
Washington DC, while the other set (Paul), who made a trip in an aeroplane via
the West Coast, Guam, Bombay, Rome,... arrived back in Washington expecting an
older Pierre, who remained in his room, as in the relativity fables. But, Paul
found Pierre to be younger than him! What does it say for the theory? This
experimental result should have been an eye-opener which revealed that the
time dilation was not decided by relative velocity, but it wasn't to be. This
unexpected turn made the experimenters to invent an imaginary frame-fairy
watching all this from an encompassing space, much like a privileged frame. In
effect, \emph{they used a limited version of Lorentz's ether relativity to
explain the experimental results, but attributed the credit to Einstein's
theory}.

Everybody wasn't convinced. Louis Essen, credited as the inventor of the
practical Caesium atomic clock at the National Physical Laboratory in England
(1955), had been involved in many precision tests of the theories of
relativity. He was the force behind the current definition of the standard
second, in terms of the atomic transition frequency in Caesium. His method of
deducing the velocity of light from the measurement of its frequency and
wavelength in a resonator is the modern method for fixing the velocity of
light.\footnote{The most precise value of the speed of light is not really a
measured of the `speed' from the propagation of light. It is the product of
two precision measurements of the frequency and wavelength ($c=f\lambda$) of a
standing wave in a cavity resonator.} In the early seventies he turned a
critique of the theory, arguing that it was just a transformation theory on
units and not a physical theory. He was critical of reducing the two
fundamental units of length and time into just one, with the speed of light
`c' as the conversion factor. His thoughts on relativity were expressed in the
tiny monograph, `The Special Theory of Relativity: a Critical
Analysis'~\cite{Essen-critique}. Soon, he was excluded from publishing his
criticisms on these topics, especially on the transported Caesium atomic
clocks experiments.

\section{Simultaneity: Einstein vs. Bergson}

Bergson's brief comments at the Sorbonne during Einstein's visit were mostly
focused on the notion of simultaneity, after a brief mention of multitude of
time in the theory. The concept of simultaneity is of central importance in
Special Relativity, being an essential prerequisite for defining
synchronization of clocks and the measurement of time. According to Einstein,
the common sense notion of simultaneity was defective and could not be
maintained in a physical theory. Events happening in two spatial locations,
appearing simultaneous to one observer because he perceive them at the same
time on his clock, will not be in general simultaneous to another observer who
is moving relative to the first. The time of an event can be reliably
attributed only by consulting a clock adjacent to the event. Then, one has to
specify a method to synchronize such clocks at different locations, to build a
consistent physical theory. That was how the constancy of the velocity of
light in all inertial frames became the pivotal point of relativity. If all
clocks are synchronized in the rest frame of the clocks using light, they are
all properly synchronized, independent of any common uniform velocity they may
have. In other words, they are synchronized as if they are all at rest.

The common sense notion of simultaneity is linked to the notions of a
universal time, and also the psychologically `felt' time and duration. Bergson
referred to a temporal `cut' of the universe of events, invoking absolute
simultaneity as a legitimate concept. He was objecting to its denial citing
the obvious point that `news' of two events from a distance, brought by the
messenger light, as simultaneous to one observer was not simultaneous to
another who intercepted the messenger signals at another location because of
motion. Bergson was trying to reconcile the common sense notion and the role
of the concept in the physical theory. Also, when one says that the event is
labelled with a nearby clock, one will have to make precise the notion
`nearby'. He asserted that these considerations at a microscopic level need
further clarification with more precision (`we are more Einsteinian than you,
Mr. Einstein'~\cite{Paris-debate}). As before, \emph{the main point of
contention was not whether simultaneity for one could become succession for
another observer in natural phenomena, but whether that happens in Einstein's
theory with its peculiar postulate}. That it happens in Lorentz's theory is
discussed by Bergson himself (in D-et-S, chapter~1), without ambiguity. Does
it happen in Einstein's Special Relativity?

We know that the sound of two bells originating in two locations A and B,
equally distant to our left and right, will be heard simultaneously, while
standing motionless at the midpoint. Another person co-located with us when
the bells were sounded, but moving relative to us, will hear them one after
the other, i.e., in succession, because he has moved away from the midpoint.
The Galilean nature of the propagation of sound waves, related to the
existence of the medium as a privileged frame, is the reason for this familiar fact.

During his intervention, Bergson described clearly how simultaneity in one
system S became succession in another, S', that was moving relative to S. He
correctly identified the relation between simultaneity and the synchronization
of clocks using light signals. (Poincar\'{e} had discussed these concepts with
unambiguous clarity before 1905~\cite{Poin-sync}). He did not object to these
conclusions in the theory of relativity. Bergson was simply insisting on the
need to `find to what extent it renounces intuition, to what extent it remains
attached to it'. He concluded by raising the hope that the theory of
relativity had nothing incompatible with the ideas of common sense. But he did
not elaborate on what was required to achieve this compatibility.

Einstein's reply at Sorbonne was brief~\cite{Paris-debate}. He summarized what
he thought were the points raised by Bergson by saying,

\begin{quote}
The question is therefore the following: Is the time of the philosopher the
same as the one of the physicist?
\end{quote}

And finally answered,

\begin{quote}
There is not a time of philosophers; there is only a psychological time
different from the time of the physicist.
\end{quote}

The transcripts of this brief and restricted encounter are inadequate to see
the really serious mature analysis by Bergson. For that one has to dwell into
D-et-S. There, Bergson analysed Einstein's own example for establishing the
relativity of simultaneity, and showed that the entire argument was inconsistent.

\subsection{Einstein's train and platform scenario}

We start with Einstein's own example, elaborated in his `Relativity: The
Special and General Theory'~\cite{Eins-book}.%

\begin{figure}[ptb]%
\centering
\includegraphics[
width=2.8837in
]%
{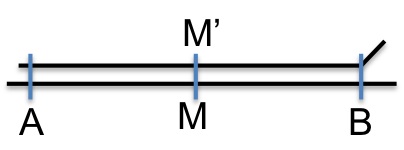}%
\caption{Einstein's example reference systems for discussing the relativity of
simultaneity. The train and the railway platform are in relative motion, with
`midpoint' observers M' and M.}%
\label{Eins-train}%
\end{figure}

\begin{quote}
We suppose a very long train travelling along the rails with
the constant velocity v and in the direction indicated in the figure
(\ref{Eins-train}). People travelling in this train will with advantage use
the train as a rigid reference-body (co-ordinate system); they regard all
events in reference to the train. Then every event which takes place along the
line also takes place at a particular point of the train. Also the definition
of simultaneity can be given relative to the train in exactly the same way as
with respect to the embankment. As a natural consequence, however, the
following question arises:

Are two events (e.g. the two strokes of lightning A and B) which are
simultaneous with reference to the railway embankment also simultaneous
relatively to the train?

\end{quote}

\begin{figure}[h]%
\centering
\includegraphics[
height=0.9061in,
width=4.2379in
]%
{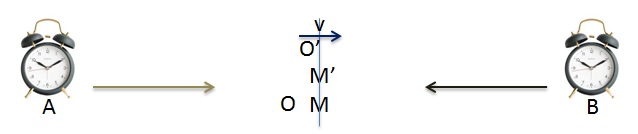}%
\caption{The situation regarding simultaneity with Galilean waves as the
messenger of events that are spatially separated. The lack of absolute
simultaneity is, however, only apparent because the Doppler effect.}%
\label{sound}%
\end{figure}

The problem is clearly stated. We will now try to answer the query in two
steps; first, under the assumption that the relative velocity of the
`messenger waves' that carries the information of the events is Galilean, and
second, that is an invariant, as Einstein postulated for light. Consider two
observers O and O' in relative motion. O is stationary and at the midpoint M
relative to the sources of the Galilean waves, say sound from two alarm bells
(see figure \ref{sound}). Therefore, O hears the bells simultaneously, say at
time $t=T_{0}$. Conversely, if O hears the bells simultaneously, then it is
deduced that they were sounded at the same time ($t=T_{0}-d/c_{s}$) at A and
B. The observer O', who coincides with O at M'=M when the bells were sounded,
is in motion towards B at the relative speed $v$. In his rest
frame, the waves are arriving at relative speeds $c_{s}+v$ from B and
$c_{s}-v$ from A. Therefore, O' will hear the bell from B first, at $t'_1=L/(c_s+v)$ and that from A later, at $t'_2=L/(c_s-v)$, and the
events would be judged as not simultaneous. Thus, the
unambiguous answer to Einstein's query for the case of Galilean waves is that
different observers perceive the order of events differently, depending on
their state of motion. Events that the stationary observer perceives as
simultaneous will not be simultaneous for the observer in motion. Identical
conclusion is reached from the rest frame of the stationary person as well;
the moving person moves away from the midpoint and therefore the sounds of
bells reach him in succession from the locations B and A. There is no reciprocity.%

We note an important additional point that is obvious in the case of sound.
Not only that the observer O' perceives the bells in succession, he would also
perceive them shifted relatively in pitch; the one from B at a higher pitch
than the one from A, due to the Doppler effect. Hence, he has a measure of his
real motion and velocity relative to air. Therefore, absolute simultaneity is
recoverable when the messenger of the spatially separated events is Galilean.

Now we analyze the same situation \emph{under the assumption that the speed of
light is the same in all inertial frames, independent of their velocity}, as
Einstein had assumed. Due to this cardinal difference from familiar Galilean
waves like sound, we expect a different conclusion regarding simultaneity when
light is the messenger, instead of sound. Then, we will examine the analysis
and answers given by Einstein and Bergson. To avoid the prejudice of
inadvertently choosing one frame over the other as privileged, we deal with
the question rigorously from two identical reference systems, both of which
are trains in relative motion, instead of a platform and a train.

\begin{figure}[ptb]%
\centering
\includegraphics[
height=1.3311in,
width=4.3837in
]%
{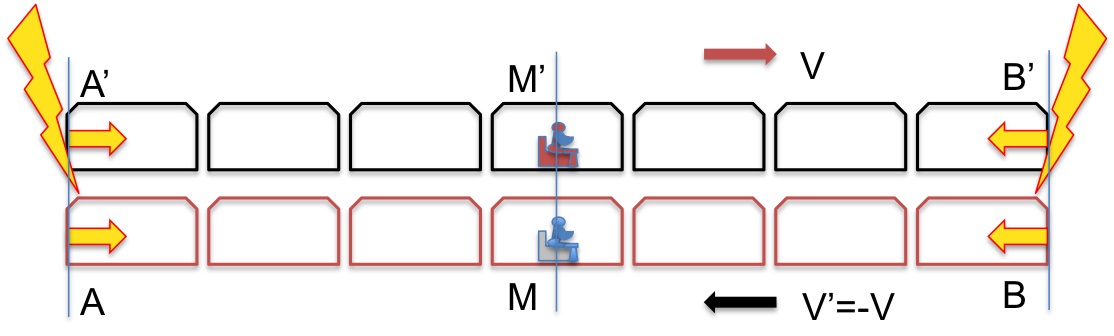}%
\caption{Reference frames S and S' with relative motion at speed v. Lightning
strikes the end points A and B (and A' and B') when the midpoints M-M' of
systems coincide.}%
\label{Rel-Trains}%
\end{figure}

In the rest frame of the train S', all its clocks read the same time and
observer at M' at the midpoint is at rest. Relative to S', the train S is
moving. But that is irrelevant for what the observer of S' experiences in S'.
After lightning strikes A' and B', light travels at equal speed from the
equidistant points A' and B' towards M' (the relative velocity of the source
does not affect the relative velocity of light). Then, the light flashes will
arrive \underline{simultaneously at M'} and the \emph{observer perceives the
events as simultaneous}. Now, we consider the observer in S. He is at rest in
S. Relative to S, other frames like S' are moving, but that is irrelevant for
what the observer of S experiences in S. Light pulses travel at equal speed
from equidistant points A and B towards the midpoint M. They will reach
\underline{simultaneously at M}, and the observer there will perceive the lightning at
A and B as simultaneous. So, we conclude that the \emph{experiences of
observers in S and S', in their own rest frames, are identical, assuming the
universal speed of light}. Let us call this the unambiguous conclusion U.

We note that if light propagated similar to sound and if its relative speed
was \emph{not a universal constant}, $c$ in one system (say in S') and unequal
($c\pm v$) from A and B in S, the events would have been seen as one after the
other, in succession, in S. This is the situation in Lorentz's ether
relativity. In a frame immobile in the ether, light travels at identical speed
in all directions, whereas in frames that are moving in ether, the speed of
light is not isotropic. Thus, light pulses originated from the equidistant
points A and B arrive one after the other. This can also be interpreted from
the immobile (ether) frame with the same conclusion; as light travels at speed
c, the frame that is moving relative to the ether is rushing towards one pulse
and moving away from the other, hence the appearance of succession. The nature
of propagation of the waves determines the nature of simultaneity.

What was Einstein's answer, which is also today's physicists' answer?
Continuing with what Einstein wrote,

\begin{quote}
Are two events (e.g. the two strokes of lightning A and B) which are
simultaneous with reference to the railway embankment also simultaneous
relatively to the train? We shall show directly that the answer must be in the negative.

When we say that the lightning strokes A and B are simultaneous with respect
to the embankment, we mean: the rays of light emitted at the places A and B,
where the lightning occurs, meet each other at the mid-point M of the length
A--B of the embankment. But the events A and B also correspond to positions A
and B on the train. Let M' be the mid-point of the distance A--B on the
travelling train. Just when the flashes of lightning occur (as judged from the
embankment), this point M' naturally coincides with the point M, but it moves
towards the right in the diagram with the velocity v of the train. If an
observer sitting in the position M' in the train did not possess this
velocity, then he would remain permanently at M, and the light rays emitted by
the flashes of lightning A and B would reach him simultaneously, i.e. they
would meet just where he is situated. Now in reality (considered with
reference to the railway embankment) he is hastening towards the beam of light
coming from B, whilst he is riding on ahead of the beam of light coming from
A. Hence the observer will see the beam of light emitted from B earlier than
he will see that emitted from A. Observers who take the railway train as their
reference-body must therefore come to the conclusion that the lightning flash
B took place earlier than the lightning flash A. We thus arrive at the
important result:

Events which are simultaneous with reference to the embankment are not
simultaneous with respect to the train, and vice versa (relativity of simultaneity).
\end{quote}

Einstein's conclusion was identical to the one we arrived at in the case of
Galilean waves (sound)! Decades of reading these passages by generations have
not spotted why the conclusion with light ended up in preferential
simultaneity of the lightning flashes for the observer on the embankment and
events in succession for the observer in the train (as it would happen with
sound waves) in a theory in which both observers are equivalent; they have the
same speed of light, and either can consider himself as at rest and the other
one as moving. Where is the gap of logic in Einstein's analysis that led to a
different inference from the conclusion U we arrived at earlier? If Einstein
had started with the train as the reference, he would have reached the
conclusion that the events were simultaneous in the train, but not on the
embankment that moved relative to the immobile train. That is what follows
from assuming the same relative velocity light in either frame. It is clear
that Einstein would have been saved from this error of inadvertent Galilean
preference for the embankment as `really stationary' if he had considered two
trains in relative motion.\footnote{This is not the only instance when
Einstein's usually clear writing generated long lasting confusion, which was
never properly clarified. The question whether a rotating disc has the ratio
of its circumference to radius as $2\pi$ or different due to length
contraction, was answered by Einstein as $>2\pi$ and by most others as $<2\pi
$. The issue is continued to be written about. What is your answer?} That the
experience of simultaneity will be identical is clear from our discussion with
the reference frames of two trains. A few like Bergson did find a problem with
Einstein's logic and expressed disagreement.

\emph{What Einstein describes is not what the observer in the train
experiences, but what the person on the platform, who considers M' as moving,
confers on the moving observer as his experience}. However, a crucial detail
of physics is missing even in that; the train of thoughts of the observer on
the platform should have moved further in Einstein's analysis as,

\begin{quote}
Now in reality (considered with reference to the railway embankment) he is
hastening towards the beam of light coming from B, whilst he is riding on
ahead of the beam of light coming from A. Hence the observer will see the beam
of light emitted from B earlier than he will see that emitted from A.
\emph{The light from B will be more blue and the light from A more red,
compared to the light pulses seen by the stationary observer, due to the
Doppler effect}...
\end{quote}

Thus, even if one goes with the way Einstein analyses the scenario, the
Doppler effect will allow the detection of true motion and then simultaneity
can easily be regained by correcting for the shift. Einstein's imagined lack
of simultaneity is Galilean and apparent, and not fundamental, because the
observer who experiences the events in succession will also see the signals
differing in their frequency content due to the Doppler shift (`imagined',
because it is not consistent with the assumption of an invariant relative
speed of light).

\subsection{Simultaneity in D-et-S}

Armed with this clarity, we can now examine Bergson's analysis of simultaneity
in D-et-S (Chapter 4). Bergson covers both physical and philosophical issues
in his comments. We discuss here only the physical and logical aspects. He
starts by quoting Einstein's scenario of the train and the platform and his
conclusion of relativity of simultaneity verbatim (the passages quoted above).
Then he writes,

\begin{quote}
This passage enables us to catch on the wing an ambiguity that has been the
cause of a good many misunderstandings...we must not forget that the train and
the track are in a state of reciprocal motion... Let us now emit our two
flashes of lightning. The points from which they set out no more belong to the
ground than to the train; the waves advance independently of the motion of
their source.

It then becomes evident at once the two systems are interchangeable, and that
exactly the same thing will occur at M' as at the corresponding point M. If M
is at the middle of AB and if it is at M we perceive a simultaneity on the
track, it is at M', the middle of B'A', we shall perceive this same
simultaneity in the train...what is simultaneity with respect to the track is
simultaneity with respect to the train.
\end{quote}

So, Bergson reached the conclusion that contradicted Einstein's incorrect one. Remarkably,
it is the same as our conclusion U. Bergson showed that the asymmetrical
conclusion on simultaneity that Einstein arrived at was merely a result of the
preferential choice of the frame of the platform as `stationary' and that of
the train as `moving'. If we had started with the train as the rest frame and
platform as moving, we would come to the opposite conclusion. Since both the
frames are equivalent, their simultaneities are also identical. The irony of
being misunderstood and judged wrong on this aspect by generations is indeed
tragic. Tragic for both Bergson, and for generations of the entire physics community.

We have seen that on both counts of multitude of times and the relativity of
simultaneity in Special Relativity, Bergson was on solid logical and technical
ground, as it can be verified easily. We also stress again the important fact
about the Doppler asymmetry, which is never brought up in such discussions. It
is obvious that an observer M' `hastening towards the beam of light coming
from B, whilst riding on ahead of the beam of light coming from A' will see
light from B blue-shifted and that from A red-shifted, by approximately the
fraction $\pm v/c$. Thus, experience of succession of the two events is
associated with the Doppler asymmetry and unequal relative velocities of
light. In modern times, a spectacular example is provided by what is called
the `dipole asymmetry' in the cosmic microwave background radiation, by which
we know that we are moving in this universe at a velocity of about 369 km/s,
accurate to a m/s.%

\begin{figure}[ptb]%
\centering
\includegraphics[
height=1.8389in,
width=2.9326in
]%
{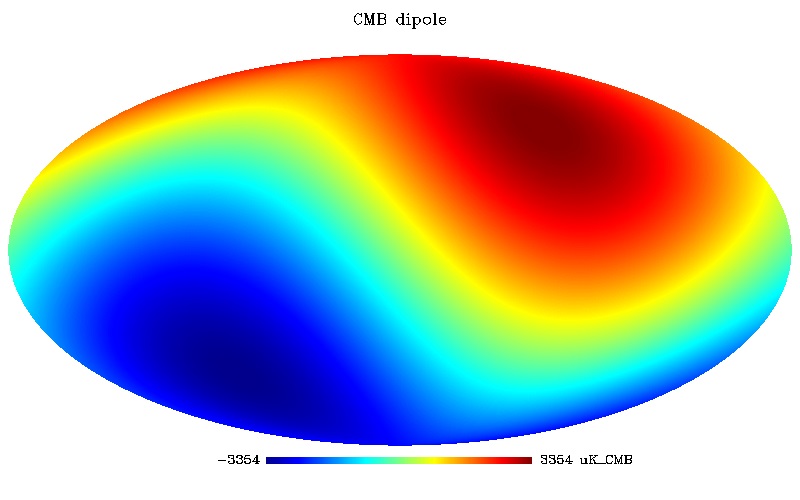}%
\caption{The dipole anisotropy in the temperature of the cosmic microwave
background, as measured from the earth, indicating precisely the velocity of
the earth relative to the isotropic universe. The average temperature
($\approx2.73$ K) is Doppler shifted up (red in colour) by $v/c$ in the
direction of motion and down, in the opposite direction. The precision of the
measurements are now good enough to measure the velocity relative to the
universal frame accurate to about a m/s! Seen as a cosmic clock, this is the
same as the universal simultaneity appearing as a succession.}%
\label{Dipole}%
\end{figure}

\section{The Hadamard Catastrophe}

In one session at the Coll\`{e}ge de France, a question was posed by Prof.
Jacques Hadamard on one of the simplest solutions of Einstein's General Theory
of Relativity, which is the theory of gravitational phenomena. Hadamard was
the professor of celestial mechanics at the Coll\`{e}ge de France. The
solution in question was found by Karl Schwarzchild in 1916, within a year of
the publication of Einstein's theory~\cite{Schwarz}. It represented the
gravitational field outside a spherical mass, like the earth or the sun. The
solution, however had a noticeable mathematical issue. It had a terms that
could go to zero or infinity as the distance from the centre of the mass
reached a critical value. Could it become infinite in some situation in
nature, when mass is very high or when the object is very compact?

Hadamard's question was about what one calls a black hole and its event
horizon in the modern terminology. Einstein was concerned about the issue. He
called it the `Hadamard Catastrophe'. The physics of stellar evolution was not
known properly in 1922 and it was not decidable whether the embarrassing
situation would occur in the case of sufficiently massive stars. S.
Chandrasekhar's discovery of the continued gravitational collapse of stars
beyond a critical mass was still about a decade away~\cite{Chandra}. Einstein
had done some calculations that indicated that the gravitational time dilation
would become so large as to freeze time before any physical quantity becomes
infinite. He wondered whether the `energy of matter transforms into energy of
space'. Einstein did not want to speculate further. And Hadamard was
apparently satisfied that some other physical process would likely intervene
before an infinity appeared in the horizon.

There is a description of the Schwarzchild solution that avoids the infinity,
but not the freezing of time. Curiously, one of the participants in the Paris
sessions, Painlev\'{e}, had already found its mathematical form in
1921~\cite{Painleve}. Painlev\'{e}'s coordinates turn out to be just a
Galilean transformation to the free-fall frames, similar to the one found by
Langevin for rotating frames, discussed in the next
section. Thus, the black hole solution in General
Relativity is not dependent on Special Relativity and its features. Many
conceptual problems related to the event horizon and time are discussed even
more vigorously in the modern times, in the context of astrophysical as well
as imaginary black holes.

\section{Sagnac's Proof for Ether}

There was one physicist in the audience at the Coll\`{e}ge de France who could
have contributed decisively to the debate about the nature of the theory of
relativity, with his \emph{experimental results}. However, Georges Sagnac's
intervention was not effective and was not even mentioned by people like
Nordmann. This detail is from the memoirs of the Nobel laureate Alfred
Kastler, who was a first year student at the Ecole Normale Sup\'{e}rieure in
1922~\cite{Kastler}. As a student attending Langevin's lectures, he got access
to the events at the Coll\`{e}ge de France. He writes from the `unforgettable
memory of those sessions',

\begin{quote}
Allow me to recall a memory from my youth: I was a first year student at the
Ecole Normale Sup\'{e}rieure at Ulm street, when Paul Langevin had the bold
idea in 1922 to invite Einstein in Paris. The undertaking at that time was not
without risk because Einstein was then professor in Berlin and anti-German
demonstrations could be feared. However all went fine. Einstein started to
give a talk for a general audience about the idea of relativity in the packed
big auditorium of the Coll\`{e}ge de France...

There were dramatic moments too. For instance when old Sagnac, the inventor of
the ingenious interferometer, gave vent to his anger against the theory of
relativity, on which he put all the blame. The only way was to let the storm
die down.\footnote{This incident of `violent intervention' is mentioned by the
physicist and writer J-P. Pomey as well, who was present at the session (Les
conf\'{e}rences d'Einstein au Coll\`{e}ge de France, Le Producteur (1922)
\textbf{8}, 201-206). Despite its intensity, Sagnac's outburst was ignored and
quickly forgotten.} And there were difficult moments too: when great
mathematician Paul Painlev\'{e} talking about the adventure of the two friends
-- the one who stayed in a place and the one who left by train and came back
-- refused obstinately to understand why the latter had remained younger than
the first one. One must say that the trivial interpretation given about this
effect is not satisfactory, because one forgets to mention that the traveller,
while turning back, is subjected to a considerable acceleration, and that to
be treated in depth, the problem must be studied from the point of view of
general relativity.
\end{quote}

There are references to two fundamental issues here. The one on the twin
paradox, debated by Painlev\'{e}, we have already discussed. It is mentioned
here again to stress the point that the paradox has no satisfactory solution,
according to Kastler, within Special Relativity. His opinion was that the
problem must be studied within the theory of General Relativity, and hence
involves gravity, as Einstein had done. (This view has been dismissed by many
modern authors, citing irrelevant reasons -- it is unlikely that they have
read Einstein's reasoning and the simple General Relativity treatment of this
problem -- see appendix I).

Let us turn to the other issue raised by Saganc that is extremely important
for the history of physics and for the physics of relativity. There is no
detail on what exactly Sagnac said, but the context is obvious. Georges Sagnac
was well known for his research in X-rays and optics. In 1913 he used a
closed-loop optical interferometer to measure the effect of rotation through
the `stationary ether'. He presented the positive results, `the proof of the
reality of the luminiferous aether by the experiment with a rotating
interferometer', in several publications~\cite{Sagnac1,Sagnac2}. He considered
his result as the decisive disproof of the etherless Special Theory of Relativity.

It is obvious that the import of Sagnac's results was largely ignored. That
was the reason for the `angry excitement of old Sagnac' that Kastler wrote
about (Sagnac was only about 50 years old in 1922, but perhaps was perceived
to be older. Details on Sagnac's scientific career and work was researched and
published recently by O. Darrigol~\cite{Darrigol-Sagnac}).

A. A. Michelson had published a similar idea in 1904, to detect earth's
rotational motion relative to the ether, but managed to do an experiment only
in 1924, with H. G. Gale~\cite{Michelson1927,Michel-rot,Michel-Gale}. The idea
of similar `first order' experiments in the context of the ether dates back
further; see the discussions by Oliver Lodge in 1893 and 1897~\cite{Lodge1897}. Michelson concluded from their heroic experiment, with a partially evacuated
kilometer-size closed rectangular interferometer, that the slow rotation of
the earth relative to the stationary ether was detected. As stated in the
abstract of their paper, they found that the \emph{beam traversing the
	rectangle in the counter-clockwise direction was retarded} (that is the
direction of the rotation of the earth). Michelson was aware of Sagnac's
experiment, but apparently believed that he could possibly get a different result because of the possibility that he massive earth might drag the ether with it, near its surface~\cite{Eins-papers-Epstein}. Somehow, Langevin and others, who commented extensively on
Sagnac's experiment, have not analyzed the Michelson-Gale experiment explicitly.

Sagnac effect, as it is called now, is one of the most important technical
tools to measure rotations optically, while being in the rotating frame. It is
the basis of the fibre-optical gyro; one may call it the `absolute rotation'
detector. Conceptually, the Sagnac effect is in the class of the Foucault
effect, the turning of the plane of a simple pendulum in a rotating frame like
the earth.%

\begin{figure}[ptb]%
\centering
\includegraphics[
height=1.3187in,
width=4.3215in
]%
{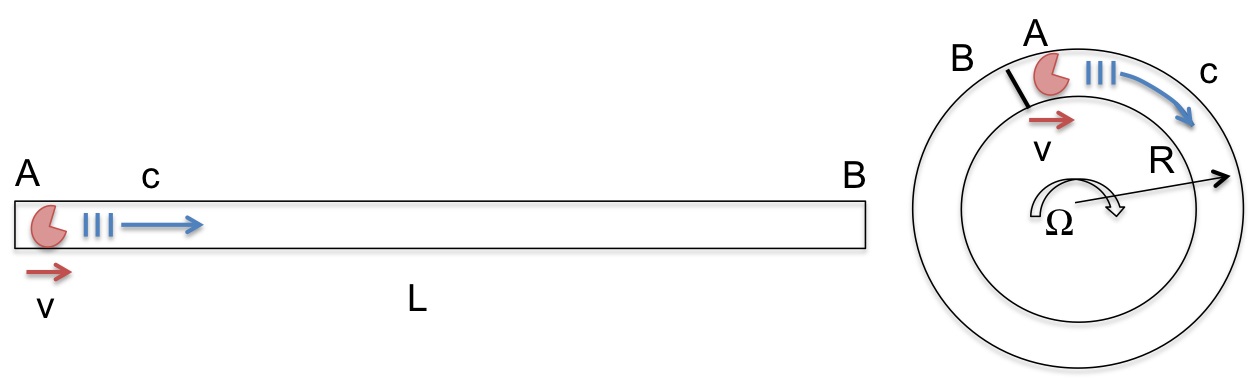}%
\caption{The equivalence of the relative durations the wave takes to propagate
a relative distance $L$, while the frame is moving at the velocity v, for the
linear case (left) and for the looped circular case (right). Here, $2\pi R=L$,
and $R\Omega=v$. The experiment on the left is not feasible, because the waves
would be far from the observer, at the end B, where another synchronized clock
is required. The one on the right is the Sagnac experiment.}%
\label{Sagac-logic}%
\end{figure}

The essence of Sagnac's experiment can be stated in a simple way, either from
the rest frame of interferometer or from the `ether frame' relative to which
the interferometer rotates (Sagnac's own reasoning was more involved, and was
tied to his own theory of the ether~\cite{Darrigol-Sagnac}). Imagine the ether
medium, relative to which light travels at velocity $c$, taking duration $T$
to travel a fixed distance $L$. If one moves in the direction of the light
waves at velocity $v$, chasing light through the stationary ether, light waves
will be moving at a lesser relative velocity, at $c-v$, just as the relative
velocity of sound waves changes for an observer who moves through stationary
air medium. Then the duration of propagation over the same distance will be
longer. If we reverse the direction of the light waves, by reflection as in
Michelson's experiment, the relative velocity increases, $c+v$, duration
decreases and nearly cancels away the increase during the up-trip. However,
instead of reflecting and reversing the path, if we arrange to loop the path,
by identifying the end points of the total path, the frame is moving in the
same direction through the medium anywhere along the path, and the
cancellation is avoided! Thinking about the case of sound, one can see that
this gives the same result as measuring the relative velocity along a straight
path, $c-v$, if we rotate in one direction. If light is sent opposite to the
rotation, the relative velocity is $c+v$. The two can be combined using two
identical waves going clockwise and counter-clockwise, while the rotation is
only in one sense; then the time taken to cover the same loop distance $L$ (in
the rest frame of the observer) by one wave is different from the time taken
by the other,
\begin{equation}
\delta T=\frac{L}{c-v}-\frac{L}{c+v}\simeq\frac{2Lv}{c^{2}}%
\end{equation}
When the light beams are combined, this difference manifests as the phase
difference and the shift of interference fringes. (The situation is similar,
and easy to understand, for sound waves in a similar device rotating relative
to still air).

Note that we have not considered any frame outside the experimenter's rest
frame, to derive this result. The simple result can be written in a more
general form, with the same content. Since $L=2\pi R$ and $v=\Omega R$ for a
path looped into a circle, $\delta T=4A\Omega/c^{2}$, where $A=\pi R^{2}$ is
the `area' of the interferometer.

The result can also be derived from the frame of the stationary medium, as
usually done. Then, the speed of the waves is the same in both directions. So,
as the waves progress, the interferometer is rotating and the end point is
moving away from one wave and moving towards the other.\footnote{To use
Einstein's expression, `hastening towards the beam of light coming from B,
whilst riding on ahead of the beam of light coming from A'; it is easy to see
the similarity to the failure of simultaneity due to the Galilean nature of
light.} Thus, one wave will take more time, and the other less, to reach the
common end point that has moved a distance, $\delta L\simeq Lv/c$ in the
nominal round trip time of $L/c$. The total difference is $\delta
T\simeq2Lv/c^{2}$, same as the value found in the rest frame. This is the same
as the splitting of simultaneity into succession due to the Galilean nature of
the waves.

Unfortunately for Sagnac, his interpretation of the result of his experiment
remained ignored, while he is celebrated now for inventing the `Sagnac effect'
(A prominent review paper on ring laser gyros mentions that `Sagnac invented a
rotation sensor' in 1913, without mentioning that it was an experiment that
claimed the `demonstration of the luminiferous ether'). The `Sagnac term' is
invoked to account for the motion of the frame in clock comparison experiments
and in the analysis of time in GPS clocks. Michelson, though 
aware of Sagnac's work with the rotating interferometer,
did not mention Sagnac's experiment in his 1927 monograph, `Studies
in Optics'~\cite{Michelson1927}. However, a paper by L.
Silberstein in 1921 on the propagation of light in rotating systems discussed
both Sagnac's experiments and Michelson's proposed
experiment, in the context of the ether and relativity~\cite{Silberstein1921}). The intuitive
reasoning behind the experiments were similar. While Sagnac measured the
rotation of a local frame relative to the ether, Michelson measured the
rotation of the earth itself relative to the ether, according to their
interpretations. 

The analysis of Sagnac's experiment by Langevin (in 1921 and later in 1937)
was perhaps the cause of the fading away of the significance of Sagnac's
experiment for relativity, forever~\cite{Langevin-Sagnac}. We mention this
important technical remark right away. In that analysis, Langevin could hide
(unintentionally, I suppose) a pure Galilean transformation in the language of
metric in geometric General Relativity~\cite{Unni-ISI}. The
true case is easily verified by simple inspection; Galilean transformation is
\begin{equation}
t^{\prime}=t, \: x^{\prime}=x-vt
\end{equation}
This gives a transformed metrical structure from
\begin{equation}
ds^{2}=-c^{2}dt^{2}+dx^{2} \Rightarrow-c^{2}dt^{2}(1-v^{2}/c^{2}%
)+2vdxdt+dx^{2}%
\end{equation}
If the straight path is looped around, as in the Sagnac device (as in the Fig.
7), $x=R\phi$, where $\phi$ is the angle from one end point. Then the velocity
is $v=R\dot{\phi}=R\Omega$, and $x^{\prime}=R\phi-R\Omega t$ is the Galilean
transformation. Therefore,
\begin{equation}
ds^{\prime2}=-c^{2}(dt^{2})(1-R^{2} \Omega^{2}/c^{2}) +2R\Omega d\phi dt
+dR^{2}+R^{2} d\phi^{2}%
\end{equation}
So, we have the same linear Galilean transformation, now looped around in a
mathematical bending. This is the \emph{Langevin metric, which is identical to
the Galilean metric}. This is so evidently not Special Relativity or General
Relativity (the theory of gravity). A transformation that was consistent with
Special Relativity would have been the one L. Thomas found in 1926 for atomic
physics of electrons in orbits~\cite{Thomas}; a set of successive Lorentz
transformations in progressively changing directions is equivalent to a
Lorentz transformation and a pure rotation, and not to a Galilean
transformation. But then, the Sagnac effect disappears from the theory.

Therefore, any result that agrees with the Langevin metric is a direct proof
of the Galilean nature of light and could be interpreted as the proof of
existence of a privileged frame, like the ether. The language of the metric
used in Langevin's analysis has mislead us to believe, naively, that the
derivation is `general relativistic'. Sagnac was right in his logic, for
everybody to see. Thus, what Langevin did, and what people follow now, is
Galilean relativity for the case of Sagnac's result, similar to what an ether
relativist (LPTR) would do. The Galilean residual term is $Lv/c^{2}$, as we
already showed. In the unfortunate twists of history, the confused community
psyche prevailed.

Though Alfred Kastler turned his academic career and attention to the
interaction between atoms and light, eventually winning the Nobel prize in
1966 for the discovery of `optical pumping', his intellectual interests
spanned wider, including relativity, universe and gravity. (The laboratories
of the Ecole Normale Sup\'{e}rieure, where many aspects of atom-light
interactions are studied, is now called the Kastler-Brossel Laboratory (LKB),
on the rue Lhomond, crossing the rue d'Ulm). In the Einstein birth centenary
(1979) paper~\cite{Kastler}, where Kastler recounted the events of 1922, he
explored Einstein's thoughts on the gravitational origin of inertia, and Ernst
Mach's speculations~\cite{Mach} on the important topic. Kastler was obviously
not aware of the detailed published work of Dennis Sciama~\cite{Sciama} in the
1950s, `on the origin of inertia'; Kastler estimated the gravitational
potential of the matter in the universe and showed that it was of the right
order of magnitude to explain inertia as due to the gravitational interaction
with the rest of the universe.\footnote{During my visits to the
Kastler-Brossel Laboratory (after 1997), in the context of laser cooling
experiments on Helium, I remained unaware of Alfred Kastler's interest in
Mach's principle and gravity, until recently when Serge Reynaud of LKB
mentioned it and gave me the relevant papers. The link of one of those papers
to the 1922 event was a pleasant coincidence.} Where Sciama and Kastler
stopped short, is indeed the door to the true relativity!

\section{A New Beginning}

We have reached the end of our impressionist glimpses of Einstein's visit to
Paris in 1922. While focussing on the physics of the issues raised by Bergson
in the context of Einstein's theory, we examined several connected aspects, in
particular Einstein's own analysis of simultaneity and the twin paradox,
Becquerel's confusion regarding clocks in Special Relativity, and Sagnac's
1913 experiment claiming the detection of ether. A significant revelation
shown transparently is the verity of Bergson's critical view on relativity of
simultaneity in Einstein's theory. The implied Galilean nature of light then
brings forth Henri Poincar\'{e}'s clear interpretations of Lorentz's local
time, clock synchronization etc., in the context of the ether and relativity.
The missing knowledge then and now was the role of the gravity of the
matter-energy in the universe, which made brief strokes in the minds of a few,
like Sciama and Kastler. But nobody took it all the way along its logical path
to realize the profound fact that the old ether is to be replaced by the
gravity of the matter-energy of the Universe, which defines the universal
reference frame and determines the nature of dynamics. All the fundamental
theories of physics were concretised before the observational results that
indicated the vast universe and its gravity started to become
available~\cite{Hubble}. Current theories that do not incorporate this
enormous gravity are flawed. The theory of relativity that is consistent with
such a universe needs a drastic and fundamental change of the paradigm of
relativity. I call this theory \emph{Cosmic Relativity}%
~\cite{Unni-ISI,Cosrel}. Its core predictions
\cite{Unni-ISI,Unni-PIRT} have been experimentally verified. 

When the General Theory of Relativity was formulated as the relativistic
theory of gravity, the basis was the special theory, set in empty space. This
was dictated by the empirical knowledge about `space' at that time. The single
most important empirical gap that existed in the early 1920s was the lack of
knowledge about our Universe. We had only an imagined universe and its
cosmology then, with very little matter around in the very distant stars and a
few nebulae, apart from the globes of the solar system. Though the distant
stars could provide a stable reference frame to measure our motion, there were
no reliable physical arguments to give the special status to that frame as a
determining, master reference frame. Therefore, both the Special Theory and
the General Theory of Relativity were explicitly formulated with the empty
space as their background. That was the significant deviation from the
Lorentz-Poincar\'{e} theory, which had the ether as the special reference
frame. Since empty space remains empty and the same for every observer in
whatever state of motion, all observers were equivalent in Einstein's theory,
with equal status to claim the state of rest.

Einstein applied his new theory of General Relativity to describe the universe
and the gravity of its matter content already in 1917. By 1922, the new large
telescopes were starting to see much deeper into the sky, opening up the vast
extra galactic universe. In 1922, Alexander Friedmann published his ideas on
the expanding universe with matter~\cite{Friedman}. Yet, Einstein, or anybody else, did not
realize that the kinematic phenomena that happen in this universe was
happening in the large gravitational potentials of all cosmic matter and
hence, not in empty Minkowski space as Special Relativity assumed. This was
not surprising in the early days of theoretical cosmology, when Einstein
himself thought that the extent of the universe is not much larger than the
size of the galaxy. But, it was indeed logical inertia to remain oblivious to
the physical implications of the vast matter filled universe that emerged later.

\subsection*{Inertial forces: Newton, Mach and Einstein}

There was one unresolved issue though that troubled Einstein, and the problem
goes back all the way to Newton. That had to do with the appearance of new
forces without any sources when the motion is not uniform. For example, forces
are felt in an accelerating mobile, or centrifugal forces are felt by an
observer in a rotating frame. For Newton, this was evidence for the existence
of `absolute space', but he could not point to it. Newton opted for an
indirect reality of Space and Time, independent of matter, with their own
primordial existence and properties. Ernest Mach, much later, criticized
Newton for positing a non-tangible absolute space and suggested that the
inertial forces (proportional to the mass) might be due to some interaction
with the rest of the matter in the universe~\cite{Mach}.\footnote{Mach was
discussing the origin of the inertial forces which could be interpreted as the
origin of inertia, but it should not be misunderstood as the problem of the
origin of mass, the charge of gravity. Also, gravity was not invoked in Mach's
proposal presumably because it was known that the weak gravity couldn't
generate sufficient forces with the amount of distant matter known in the
cosmology of the 19th century.} Einstein sidelined the issue in Special
Relativity by considering only inertial frames that do not accelerate or
rotate. But, he had to go beyond this restriction to formulate the general
theory, which was a theory of gravitation. Einstein had enthusiastically
sought support in Mach's conjecture linking inertia and the matter in the
universe. He even elevated the conjecture to status of a Principle. Yet,
Mach's principle was left out in the final theory. There, aided by his
brilliant insight, and noticing that the inertial forces were proportional to
the mass just as gravity was, he postulated that the two were
indistinguishable -- a new principle of relativity, which is known as
Einstein's Equivalence Principle. This was a second time Einstein went ahead
converting a problem to be solved into a postulate. Thus, he sidelined the
issue again, after struggling and failing to incorporate Mach's idea into his
new theory. He retained the reality of the empty space, devoid of `stars and
the rest of the universe', as the inert background for his general theory as
well. Einstein was aware that his theory had some results that kept suggesting
to listen to Mach, but with the equality of all observers and consequent
denial of a privileged frame capable of `influence by interaction', Mach was
definitely out. Mach's stand on the inertial forces was not explicitly
discussed during the 1922 visit, though Mach's philosophical stand on certain
issues did figure in the discussion. Mach was known to be critical of the
theoretical structure of General Relativity, and Einstein thought that was
`because he was old'~\cite{Nordmann}. In any case, relativistic physics and
the theory of gravity inherited Newton's non-tangible space and time, stripped
of their absolute character and superficially fused into a space-time with a
hidden imaginary character to time (that is, when we describe space and time
together, `time' needs a multiplier $\sqrt{-1}$). Bergson commented
extensively on this peculiar \emph{spatialization of time} in D-et-S, but we
do not discuss this aspect in this article.

\subsection*{Going beyond: Cosmic Relativity}

The guiding light dawned slowly, after 1928, not too late for a rethinking on
these issues. It was the new observational cosmology of the extra-galactic
universe, after Hubble's discovery of the recessing galaxies~\cite{Hubble}.
The universe that emerged was really vast, with enough matter and mass to take
Mach seriously, even with a feeble interaction between matter. And gravity was
a well understood interaction. Unfortunately, neither Einstein nor anybody
else retraced the path of rigorous physical logic with this crucial knowledge;
the theories of relativity continue to assume that our physical space is the
non-existent empty space, and our time has no relation to the cosmic matter.
It is as if the standard physics of today considers the universe we see as a
huge make believe film set with virtual reality galaxies without gravitating
matter; the theories continue in their course set definitely in the 1920s. In
reality, there is one master frame with its material reference markers as the
galaxies, and one universal time, given by the expanding universe and its
evolutionary clock (also termed `absolute time', but only in the sense of
being defined by the single master reference frame of the material universe.
`Universal time' is the appropriate term, to avoid any confusion. This is
distinctly different from Newton's ideal and metaphysical absolute time).

The old ether was stationary and without evolution, whereas the real universe
is evolving, as evident in its expansion. Thus, the evolving universe defines
a universal time, which the ether was devoid of. The gravity of all the
matter-energy in the universe physically determines the possibilities of
relativistic dynamics. When this reality is acknowledged, one gets a new
theory of relativity, incorporating Mach's ideas, and without inconsistencies.
I have named this theory appropriately as Cosmic Relativity; it retains
naturally all the empirically correct aspects known in relativistic physics,
but it is distinctly different from Einstein's theories~\cite{Cosrel,Unni-TPU}. 
In Cosmic Relativity, dynamics and the inertial forces are \emph{derived}
from the gravitational influence of the cosmic matter. The Principle of
Relativity follows from the approximate large scale homogeneity and isotropy
of matter-energy in the universe. Motional modifications of duration and
length are really gravitational modifications, because of the modified
gravitational interaction due to motion. The propagation of light (and
gravitational waves) is Galilean, controlled by the gravity of the universe
(the gravitational potential acts like an effective universal refractive index
to limit the velocities of all propagating waves). Einstein's equations for
gravity are modified by including the cosmic gravity and the master frame as
part of the equation itself. Then, Mach is naturally integrated with Galileo,
Newton, Lorentz and Poincar\'{e}, and Einstein, retaining only the consistent
elements that constitute a satisfactory completion.

Cosmic Relativity demands and predicts that the relative velocity of light is
Galilean.\footnote{This was the earliest prediction of the theory, even before
many aspects of the theory were well understood. Galilean propagation of light
is the logical consequence of a privileged frame.} An empirically rigorous
experiment to test whether the one-way velocity of light, relative to
inertially moving reference frames, is Galilean or Einsteinian needed
innovations that avoided two spatially separated clocks and their
synchronization, yet keeping the motion inertial. When this was achieved in
our laboratory at the Tata Institute in 2005, the truth about the propagation
of light was transparently visible; light, like waves of sound, is
Galilean~\cite{Unni-ISI,Unni-PIRT}. With new technology and lasers, the
experimental task is much more simpler than what Michelson and Sagnac had to
tackle. Thus, verifiable experimental results vindicate the logical proof we
discussed, that the postulate of invariant velocity of light is inconsistent.

Cosmic Relativity firmly and surely brings back a real master frame that
determines relativity and dynamics -- the universe with its matter-energy and
its gravity, in place of the old ether. Time now has a universal reference.
But, there is real gravitational modification of duration in frames moving
relative to the master frame, apparently shattering Bergson's hope for the
single universal time. Is it possible to rebuild Bergsonian philosophy with
its universal time afresh from this new reality?  

\section*{Resurrection of Bergson's Philosophy of Time}

We conclude the article with a new insight on Bergson's philosophy of time,
with its basis on the notion of a universal time and the implied absolute
simultaneity. The reason Bergson ventured to study the theories of relativity
was to understand the conflict between his philosophy and the physical
theories, with their multitude of times and the relativity of simultaneity. In
Lorentz's theory with the insensible ether, there was the direct and
irreconcilable conflict; each observer had his own real time that depended on
the speed relative to the ether, but there was no way to sense the state of
motion. Bergson's hope was to find harmony with Einstein's theory, with its
equivalence of all observers in motion and at rest, allowing the
interpretation that the multitude of times was merely in the mathematical
content of the theory without real physical manifestations. The empirical
situation eventually proved otherwise, showing the reality of velocity
dependent multitude of durations. But, we have seen that the real reason for
the motional modification of duration could be found in the gravity of the
matter-energy in the universe. This is \emph{consistent with both the Galilean
	nature of light and the existence of a privileged master frame}. The amazing
fact is that this frame, identified as the matter-filled universe, \emph{has
	both the universal time and the physical multitude of times, coexisting
	consistently}. This is because unlike the situation in invisible and static
ether, those who move have the correct measure of their motion with the cosmic
matter and radiation as the markers of real rest, in the slowly expanding
dynamic universe. The crucial physical distinction to note is that while the
absolute nature of the old ether is truly insensible, the absoluteness of the
material universe in time and space is real and easily observable, with its
gravity manifest as the verifiable reason behind relativistic effects.

We can now assess the compatibility of Bergson's preferred view on time,
duration and simultaneity, with the factual relativity and its theory, based
on cosmic gravity (cosmic relativity). The modification of durations of a clock in motion is real,
and therefore there are a multitude of real times. Similarly, simultaneity
changes to succession for observers in motion, being linked to the Galilean
nature of the relative velocity of light. Therefore, one might think that the
Bergsonian philosophy of duration and simultaneity that needed a universal
time and an absolute division between simultaneity and succession would
crumble, as it would happen in the Lorentz-Poincar\'{e} `half-relativity'.
However, this judgement is premature. Since the factual relativity is entirely
based on the gravity of matter-energy in the universe, there is a
\emph{tangible privileged frame}, relative to which all motion is felt and
measured. No one in real motion can claim a state of rest because that will be
in conflict with what he measures in his external world filled with matter and
radiation, which are real material markers. \emph{This is a profound
	difference compared to ether relativity, in which the velocity of motion could
	not be detected by any means.} In new relativity, the modification of the
duration is related to the real measurable velocity relative to the cosmic
matter (section 5.2). Therefore, an observer always has the access to both his
proper time, measured by his clocks, and the universal absolute time measured
by the clocks at relative rest to cosmic matter! This universal time, the same
time everywhere in the universe, is operationally equivalent to the
temperature of the cosmic microwave background radiation in the expanding
universe. In fact, theoretical cosmology always worked with such a universal
time, akin to Bergson's time, but never recognized that the physics done in
this same and only universe has such a universal time.\footnote{B. Latour has
	commented on the fact that Bergson stressed the `cosmological import' of his
	notion of space and time and on the need to find consistency between
	subjective time and the time of physics (ref. \cite{Latour}). I am arguing
	that the physical universe provides this consistency naturally through its
	physical effects that can be experienced.}

Now we can see that the notion of simultaneity of spatially separated events
is also absolute, because these events are tagged to their local time which in
turn is the single universal cosmic time. The succession experienced in
perception is mere appearance, exactly as the change of simultaneity of sound
of bells for one at rest into succession for another who is moving toward one
of the bells. Since the velocity is known, simultaneity is regained, just as
the universal time is regained.

Thus, we have a surprise end to our discussion. The postulated equivalence of
the mobile and the immobile in the prevailing theory is incorrect. There is a
universal time and a well defined reference for real motion. Physical theory
of relativity, with its multitude of times and velocity dependent modification
of physical quantities, nevertheless maintains harmony with the Bergsonian
notions of a universal time and simultaneity. With the universal time
regained, Bergson's philosophy remains intact. \emph{We are able to resurrect
and complete Bergson's program in the philosophy of time, of finding
compatibility with the notions of time and duration in relativistic physics }
\cite{Unni-Bergsonian}. Then, the time of the physicist and the time of the
philosopher, and indeed the time of common intuition, all have the same basis
and coexist without any logical or conceptual conflicts.

\section*{Acknowledgements}

Many sections of this article were refined with the help of Martine Armand,
whose translations and readings, while appreciating the angel's share, were
important in going into details of the discussions and academic interaction
during the 1922 visit. Continued help on many occasions from Mich\`{e}le
Leduc, my academic host for several years at the Kastler-Brossel Laboratory of
the Ecole Normale Sup\'{e}rieure (LKB, ENS, Paris), was always a supporting
factor. I am thankful to Serge Reynaud of LKB, ENS, who alerted me to Alfred
Kastler's long term interest in Mach's hypothesis and provided the relevant articles.

\section*{Appendices: Background to This Section}

In the main text we have discussed how the position taken by Einstein and his
followers was not fully justified. We have already proved transparently how
Einstein's analysis of the central aspect of simultaneity relative to two
frames in relative motion was in conflict with his own postulate of the
invariance of the velocity of light, confirming Bergson's comments on the
issue. 

The crucial insight that makes a drastic revision in our understanding of the
theory of relativity is the definite knowledge of our universe, which was
lacking in 1905-1915, or even in 1922. Thus Einstein's theories of relativity
are built on the explicit and necessary assumption that space is empty --
devoid of any matter in general, like Newton's space. With no matter or
material markers, empty space remains isotropic to an observer who moves in
space. That is how Special Relativity has the invariance for the velocity of
light and General Relativity has the general coordinate invariance. In the
present paradigm, the universe is treated as just one of the many possible
solutions of the Einstein's equations of gravity in General Relativity, and
not as an \textit{a priori} premise for all physical phenomena.

But this is obviously deeply flawed. The theories and equations of physics are
operative in this universe, in the presence of all the matter-energy and its
gravity. All tests of relativity and indeed all physical processes and
dynamics happen in the presence of this gravity, but this presence is not part
of the present theories! Even a cursory logical inspection reveals that we are
seriously off the track and the correct theory should acknowledge and
incorporate the privileged frame of the matter-marked universe, and a
universal time in the ever expanding universe (also available as the
monotonically decreasing temperature of the cosmic radiation background). The
observer in real motion relative to this matter experiences a different world
and physics compared to an observer who is at rest relative to this matter,
due to the difference in the gravitational potentials in these two situations.
This physical fact should be the basis of the correct theory of relativity, as
achieved in the theory of `Cosmic Relativity'~\cite{Unni-ISI,Cosrel}. This
unified theory of relativity and dynamics replaces the prevailing theory, and
with that Einstein's General Relativity also is modified to include the
gravity of the universe as part of the fundamental equation itself. There is
no general coordinate invariance and the world of physics is factually
Galilean and Machian, with all the empirically seen relativistic effects
correctly predicated as gravitational effects of the matter in the universe. 

We examine the empirically verifiable physical links between duration, motion
and gravity in the appendices. In particular, I present Einstein' own
resolution, in 1918, of the twin paradox \cite{Eins-twin}, which should be an
eye opener to those who are unfamiliar with that paper. 

\section*{Appendix I: \newline The Clocks of Twins and the Triplets}

The traditional debate regarding the `relativistic twins' was about who ages
more, Pierre or Paul. We know empirically that some transported clocks, like
Paul, age less. We also know that some transported clocks, age more when they
come back after their trip! A clock that is inertially transported, after
synchronizing, from Mumbai to Paris by surface transport will actually age
faster, and not slower \cite{Hafele-Keating}. To check this, one can
accumulate the durations by transporting it further across the Atlantic and
Pacific oceans, and back to Mumbai. The travelled clock would be about 100 ns
\emph{older} than the stationary clock in Mumbai, which never left the room.
The reason for this cannot be found within Special Relativity. But in a theory
with an absolute master frame, this is easy to understand, The `frame at
rest', in which Pierre resides, is really moving relative to the master frame
and we should use that velocity to calculate the total real time dilation, and
not the relative velocities of Special Relativity. However, this is not what
we want to discuss here, because the experimental results are
known~\cite{Unni-TPU}, and such a calculation can be done by anybody to verify
the facts.

We go back to the question addressed by Bergson: how does one decide that one
of the travellers ages less in the observer-symmetric Special Theory of
Relativity (STR)? First, suppose one knew about only STR. Now we have three
characters of the same age, Pierre, Paul and Pauline. Paul and Pauline are in
their spaceships. Pierre, who is on a large space station, perhaps like a
planet, is moving away relative to the other two. The large velocity of
separation is $v$. Their clocks are synchronized before they separated. Who
ages faster? We see that we are unable to answer this question, because we
formulated the question correctly! Try answering it, in place of Becquerel and
the many Einsteinian relativists through times, and we are immediately stuck,
because Special Relativity is a symmetric theory. Paul and Pauline think that
Pierre should age slower because they see his moving away, while they are at
rest in their spaceships. Pierre thinks that he should age faster than Paul
and Pauline who are moving away together. In a theory with a master frame,
like the universe and its matter, there is no confusion. Those who move
relative to the master frame, age slower. But in STR, at the end of half the
story, a few years of separation, there is no definite answer to our question.
Then the adherent of STR listening to the paradox fables would ask,
\textquotedblleft then what happened?", looking to break the symmetry.

Here, Bergson's questions as a philosopher, also reflecting common intuition,
take relevance. Shouldn't a theory be able to give the answer at all times?
Why does the theory need to know what happens in the scenario later, to say
something about the \emph{lived durations}? Surely, one of the clocks should
be slower than the other, if such a physical effect is really there. Which one?

At least one fact is clear -- Paul and Pauline, being side by side, age the
same way, with their clocks remaining synchronized. The clocks of the triplets
have some readings locally, and whatever they are, a clock's reading in one
spacecraft cannot be changed by firing a rocket in another spacecraft far
away. Nobody would agree to `spooky action at a distance' in relativity's
clocks. Since this is a very important point that is neglected in debates, let
us state it clearly. \emph{Relative time is the difference in the readings of
time in two distinct identical clocks; one quantity defined by two clocks}.
Therefore, relative duration can change only by changing one or both the
readings. If acceleration has any role in changing the proper time, it can
only affect the clock that is accelerated, a conclusion everyone would agree
with. Hence, if two clocks were synchronized at some stage, after which they
were in relative inertial motion, they would be completely equivalent is
Special Relativity, till one of them is accelerated. When such an asymmetry
occurs, only the clock that is accelerated can have a new physical effect
altering its readings. So, there are three phases for the rest frame of the
accelerated clock; two identical inertial stages of the round trip, during
which it is reciprocally equivalent to the unaccelerated clock, and one
intermediate stage of short acceleration for which we have to calculate the
time dilation of that particular clock with an appropriate theory. \emph{At
all times, the clock is at rest relative to itself}.

Pauline decides to fire a rocket thruster briefly and arranges her flight path
to be directed back to join Pierre. So, if at all there is some change to the
clocks as a result of Pauline's action, that must affect only the reading of
Pauline's clocks, and not Paul's or Pierre's. She is still near Paul, since
she needed to fire the rocket very briefly; then there is only a slight
difference between her clocks and Paul's, perhaps a few hours. Of course,
nothing would have changed unusually in Pierre's clocks, while she fired her
rocket. Since Paul and Pauline decided, using STR, that Pierre was ageing
slower and must be younger by several months by then, that conclusion doesn't
change by this difference of a few hours between the clocks carried by Paul
and Pauline. Pierre, however, continues to think that the other two must be
ageing slower. He also knows that there will only be a slight difference
between their clocks if one of them fires a rocket briefly. Who really ages
more and by how much? \emph{There is still no answer in STR}.

That is why Einstein brought in the General Theory of Relativity and
gravitation to break the symmetry and try to solve the puzzle~\cite{Eins-twin}
(commented as necessary by Nordmann and Kastler in their notes on the 1922
visit). He admitted that there
could not be differential ageing in the symmetrical theory of STR. \emph{Then the
differential ageing must be due to some other reason, to be
handled by another theory}.

\subsection*{Einstein's own resolution of the twin paradox}

The paper that Einstein wrote in 1918 to explain his resolution of the twin
paradox is important for both physics and the history of physics. Not many
physicists are aware of the paper, as evident from their continued use of
incorrect arguments for the resolution. Here is Einstein's simple General
Relativity argument, which he thought enabled him to `extricate skillfully'
from the paradox eventually~\cite{Eins-twin,Unni-twin}. First, he admitted
that the time dilation in STR, while the clocks were in uniform relative motion, was symmetric. So, Pierre reckons that Pauline
is younger by the factor $\gamma=1/(1-v^{2}/c^{2})^{1/2}$ and Pauline and Paul
would claim that Pierre is younger by the same factor. If the relative
velocity is not very high ($v<0.1c$), we may write the time dilation during
the total flight duration $2T$ as $\delta T\simeq Tv^{2}/c^{2}$ (the sign is
relative to the clock in the rest frame, indicating that the rest frame clock
is faster). While Pauline fires the rocket to reverse the trip and join
Pierre, her deceleration is $a$, for a duration $t$, such that her relative
velocity reverses, $at=2v$. In Pierre's frame, that doesn't make much
difference to her ageing, if $t$ is a relatively short duration. However, in
Pauline's rest frame, she is not moving, but just experiencing a uniform
gravitational field $g=-a$, \emph{extending over all space}, according to Einstein
Equivalence Principle. Then there should be a \emph{relative gravitational
time dilation of General Relativity between the clocks separated by the
distance} $L=vT$ equal to ${\Delta t}=tgL/c^{2}=-taL/c^{2}=-2vL/c^{2}%
=-2Tv^{2}/c^{2}$. We note the important feature that the amount of
gravitational time dilation is independent of the duration of acceleration
$t$. This is double the special relativistic time dilation and of opposite
sign, Pauline's clock running slower than Pierre's, in her frame. The total
time dilation is hence $\delta T+\Delta t=-Tv^{2}/c^{2}$, which is the same
conclusion as Pierre's; Pauline's clocks are `younger' when they meet again.

That is how Einstein solved it. Or, thought he solved it, and was confident
enough to write a paper describing the solution, with explicit statements that
the time dilations in the symmetric Special Relativity were
symmetric~\cite{Eins-twin}, a stand similar to that of Bergson! This position
contradicts most texts on the theory.\footnote{Einstein's close friend, Max
Born, wrote in his textbook, \emph{Einstein's Theory of
Relativity}\ (p. 284), ``Thus the clock paradox is due to a
false application of the special theory of relativity to a case in which the
general theory should be applied.'' However, he chose to claim that the
resolution was fully within special relativity when the twin paradox
controversy was discussed in the pages of Nature, in the 1960s (M. Born,
Nature 197, 1287 (1963)).} 

In Einstein's paper, the asymmetry of the acceleration and equivalent
gravity were handled by the equivalence principle and General Relativity. But,
he was not attentive enough, in details. There are \emph{two clocks} in the
problem and the formula is for the relative time. We just found that Pauline's
clock was not affected by her brief deceleration, as she verified by comparing
with the nearby Paul. Her firing the rocket could not have affected the clocks
with Pierre, who was $L$ distance away from her! So, there must be something
seriously missing in Einstein's solution. Even a layman can note that \emph{the
entire argument collapses if Pauline chooses to `arrest' or `freeze' the
clock during the brief period of noninertial motion}. Then her clock cannot
suffer any time dilation! Besides, take a look at the actual numbers in an
example, within the approximation he used. If $a\simeq10^{2}$ m/s for two
days, and $L\simeq10^{15}$ m, after a year's travel at $v\simeq0.1c$, all
small compared to the numbers usually used in the discussion of the paradox,
then $aL>c^{2}$; clearly the formula for the gravitational time dilation
breaks down. Further, the twin paradox arises the same way even in a scenario
where the clock readings are transferred between inertial frames (between two
spaceships travelling in opposite directions) \emph{without any physical
accelerations}. The physical clocks or the observers are not transferred or accelerated; only the information on time is transferred from 
the frame moving at the velocity $\vec{v}$ to another that is moving at the velocity $-\vec{v}$. Then there is no equivalent gravity to induce Einstein's
gravitational time dilation. Yet there is asymmetric relative time dilation. So, Einstein's resolution is not general enough
to resolve the paradox, and cannot be the genuine resolution~\cite{Unni-twin}.

However, if one rejects Einstein's resolution, by claiming an entirely special
relativistic resolution, one is rejecting Einstein's Equivalence Principle and
General Relativity. The relative time dilation of clocks in an accelerated
frame is a prediction of General Relativity. That is the serious dilemma one
has to face. As indicated earlier, the correct solution is in the paradigm of
Cosmic Relativity, in which the Einstein equivalence between clocks in an
accelerated frame and in a uniform gravitational field is not
valid~\cite{Unni-TPU}. All motional time dilations are factually gravitational time dilations in the velocity-dependent gravitational potential of the matter in the universe. 
Acceleration is not important in the time dilation problem.
It is significant that the relative time dilation of clocks in an accelerated frame is not yet tested.

\section*{Appendix II: \newline A Lesson on Time From Birds Flying Backwards}

We can briefly come back to the theme of the flock of birds seemingly going
backwards, taking it as a metaphor. In Einstein's theory, relativistic
physical effects depend on the relative velocity as the sole parameter. For
two frames in inertial motion, only their velocity of separation, or the
relative velocity, makes physical sense in empty space. However, when there is
a privileged frame and universal markers in space, the situation is different.
In the old ether relativity, it was the electromagnetic properties of the
ether that were meant to induce motional physical effects. In Cosmic
Relativity, it is the gravity of cosmic matter that is the source of all
relativistic effects. In such theories, it is the absolute velocity -- the
velocity relative to the master frame -- that determines the magnitude of the
physical effect.

Consider an inertial reference frame from which Pierre is watching the world.
Paul is moving fast to his left, and so is a flock of birds, only slower. The
birds' timing and synchronization is admirable. \textit{Pierre concludes that
Paul's clocks are slower than his}. Pierre also estimates that the faster Paul
would age slower than the birds. If they continue moving like that, birds
should age more than Paul.

Paul greets Pierre as they pass by. \textit{Paul estimates that Pierre's clock
is running slower than his}. Then he sees the flock of birds, flying
backwards. \emph{Being an adherent of Einstein's theory}, he reckons that birds in
relative inertial motion have their life and internal clocks running slower
than his. If they continue their flights, the birds should maintain their
slower ageing.

So, Einstein's theory does not give any weight to the logical inference that
Paul is indeed moving faster than the birds and therefore one expects larger
motional time dilation in Paul's clocks. As we emphasized earlier, relative
time is the difference between the readings of two independent clocks,
$t_{1}-t_{2}$. A complete physical theory should be able to provide an
expression for the modification of the rate of the individual clock, $t_{1}$
or $t_{2}$, etc., from which one can calculate the relative time dilation.
This is much like the spatial separation or the distance between two material
points. The change in distance is effected by independent changes in either
position. 

This by itself does not reveal any inconsistency in Einstein's theory in the
most common situations because an actual empirical comparison requires
bringing the clocks together. Then the usual discussions reduce to a two-way
transport of some clocks. However, think of Paul and the birds in their
continued flight. After a very long duration, Paul encounters the same flock
again, having gone around the globe in the long trip. Now he can compare the
clocks; to his shock he would find that his expectation from Einstein's theory
went wrong; the birds who were seen in relative motion actually aged faster! Special relativity is not adequate to explain the physical situation.

The correct physics of motional time dilation of clocks (as in GPS) is to be
found outside Einstein's theory. I assert from Cosmic Relativity that the
correct and most accurate expressions for the relative time (difference of
elapsed durations) of two clocks is
\begin{align}
\Delta T/T  &  =\sqrt{1-v^{2}/c^{2}}-\sqrt{1-u^{2}/c^{2}}\\
&  \simeq-(v^{2}/2c^{2}-u^{2}/2c^{2})=-\frac{(v-u)(v+u)}{2c^{2}}%
=\frac{(u-v)(v+u)}{2c^{2}}\nonumber\\
&  =-\frac{(v-u)(v-u+2u)}{2c^{2}}=\frac{(u-v)(u-v+2v)}{2c^{2}}\nonumber\\
&  =-\frac{v_{rel}^{2}}{2c^{2}}-\frac{v_{rel}u}{c^{2}}=\frac{v_{rel}^{2}%
}{2c^{2}}-\frac{v_{rel}v}{c^{2}}%
\end{align}
Here, the velocities $v$ and $u$ are relative to the cosmic frame, and
$v_{rel}=(v-u)=-(u-v)$. The asymmetry between the frames is
explicit. When the velocities are not uniform, the infinitesimal durations can
be accumulated (integrated) to get the total duration.

Positioning algorithms for the motional relativistic corrections in the Global Navigational Satellite Systems (GNSS) like the GPS were designed assuming the STR axiom of the invariant relative velocity of light. Light signals (radio-frequency) are used for the time transfer. Hence, from the measured difference of the clock times at the satellite and the receiver, $\Delta t=t_s-t_r$, the distance to a satellite was to be estimated as $d_s=c\Delta t$. In contrast, in a theory with  the universe as the master determining frame, the distance would be calculated with the Galilean relative velocity as 
\begin{equation}\label{key}
d_s=(c-\vec c \cdot \vec {v_r}/c)\Delta t\simeq c\Delta t-\vec{d_s}\cdot \vec {v_r}/c
\end{equation}
This is the equation that is factually used in GPS, and not the STR formula $d_s=c\Delta t$! Ironically, the correction is called the `Sagnac term'.

Similarly, the second order clock correction used in GPS relies on the formula (6) and not on the special relativistic formula $-v_{rel}^2/2c^2$. The `absolute velocity' of the receiver explicitly appears though it is at rest in its frame. GNSS provide ample proof for Cosmic Relativity based on gravity of the matter-energy in the universe.

\section*{Appendix III: \newline Lorentz's Local Time and the New Universal
Time}

We are familiar with the local time set according to the celestial solar
clock. This helps in adjusting the time of the day to the presence of the sun
at each locality on the rotating earth. But, this has
no deep physical significance. We do not infer the velocity of an aircraft
using the difference in local times at the start and the end of the journey.
Everybody understands this.

\begin{figure}[h]%
	\centering
	\includegraphics[
	width=4.7in
	]%
	{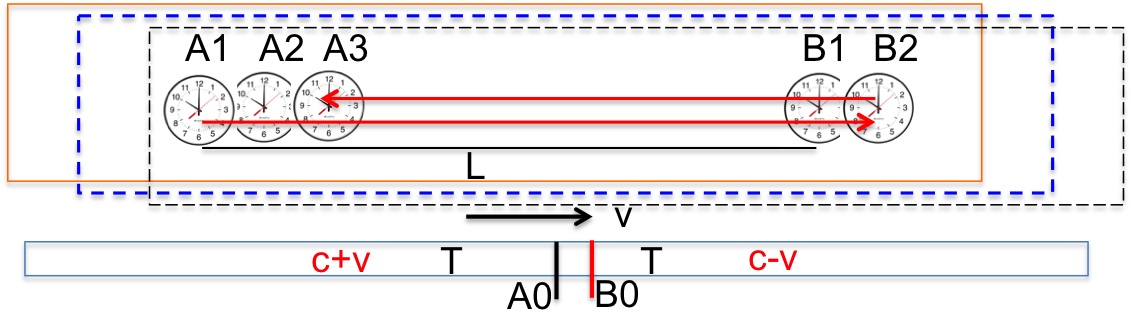}%
	\caption{Synchronization by light signals in a frame that moves at velocity
		$v$. Observers in the frame are unaware of their motion and assume that
		light signals take equal times in their up and down trips, A to B and back.
		Clock B marks zero when it receives the light pulse and A marks zero at the
		midpoint of the total duration $2T$. In reality, the clock moves to B2 while
		the signal propagates, taking more time to reach (long right arrow), and it is
		received back when the clock is at A3, taking less time in transit. Then the
		local time set at B is behind the time at A, while the observers at B \protect\underline{believe} that they have the same time as at A. If the
		separation between the clocks is $L$, the total duration $2T=2L/c$ is divided
		in the ratio $t:2T-t=(c+v):(c-v)$. Thus $B0$ is not the same as $A0$; it is
		$\delta t=Lv/c^{2}$ later than $A0$.}%
	\label{Syncing}%
\end{figure}

\begin{figure}[ptb]%
\centering
\includegraphics[
height=2.2755in,
width=2.1189in
]%
{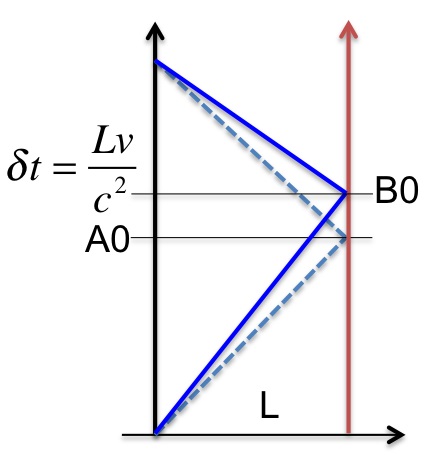}%
\caption{Clock synchronization in a space-time diagram. The dashed diagonal
lines indicate the belief about the propagation of light and the blue lines
indicate the Galilean reality, in a moving frame. The local time at B (`0' of
B) is later by $Lv/c^{2}$ relative to A0, which is at a distance L away.}%
\label{Sync-ST}%
\end{figure}

Lorentz realized that the equations of electrodynamics remain independent of
the velocity of the reference frame if the clocks at a distance $x$ from the
origin of the frame are deemed to read the local time $t'=t-vx/c^2$, where $v$
is the velocity. Lorentz assumed that this was a mathematical device that
represented the inability to detect inertial motion by local experiments, with
physical effects that were of first order in $v/c$. Poincar\'{e} correctly
explained the meaning of Lorentz's local time as related to the
synchronization of clocks done within the constraint of the principle of relativity, using the factually Galilean propagation of light (see figures
\ref{Syncing} and \ref{Sync-ST}). Poincar\'{e} remarked~\cite{Poin-sync},

\begin{quote}
I suppose that observers placed in different points set
their watches by means of optical signals; that they try to correct these
signals by the transmission time, but that, \emph{unaware of their
translational motion and thus believing that the signals travel at the same
speed in both directions}, they content themselves with crossing the
observations, by sending one signal from A to B, then another from B to
A.
\end{quote}
I have italicized the essence of the interpretation. Bergson's chapter `Half
Relativity' in D-et-S, contains a clear and mathematically accurate
description of this synchronization. Remarkably, Bergson's interpretation is
essentially the same as Poincar\'{e}'s, that the local time is the mismatch
between two clocks synchronized \emph{believing that they were at rest,
unaware of the motion} of the frame at velocity $v$. It is obvious that the
measurement of the velocity of light done using such clocks, synchronized
assuming equal velocity both ways, when it is really $c-v$ and $c+v$, will
always return equal (isotropic) velocity as the result.%

Going further, inertially moving observers, unaware of their motion, cannot
account for their own relativistic time dilation by the factor $1/\gamma
=(1-v^{2}/c^{2})^{1/2}$. Hence they believe that their clocks are reading the
same time of those at rest, which is $\gamma$ times faster than they actually
are. Thus, we get the Lorentz transformation for time, $t^{\prime2})$. For
space, there is no `synchronization'; observers in motion have a different
spatial coordinate, given by the Galilean formula, $x^{\prime}=x-vt$ and being
insensitive to their own real length contraction by the factor $1/\gamma$,
they believe that $x^{\prime}=\gamma(x-vt)$. That is the correct physical
interpretation of the Lorentz transformations.

Einstein, on the other hand, adopted the Lorentz transformations and
attributed a real physical significance to the local time; this is equivalent
to assuming that the velocity of light is independent of the velocity of the
observer. Besides being inconsistent with the measurements of genuine one-way
relative speeds \cite{Unni-ISI}, we have demonstrated the internal
inconsistency in the previous appendices on time dilation. 

\section*{Appendix IV: Galilean Black Holes of General Relativity}

Similar to the Galilean nature of the Langevin metric in rotating frames
(section 7) and the metric of an observer with proper motion velocity $v$ in
FRW universe (subsection 5.2), the Painlev\'{e} representation \cite{Painleve}
reveals the Galilean nature of the Schwarzchild black hole solution
\cite{Schwarz} of the vacuum Einstein equations ($R_{ik}=0$) of general
relativity.  Though the solution was known in 1921, it was perhaps not
explicitly discussed during the 1922 visit (section 6). The metric in the
absence of a point mass at $r=0$ is $ds^{2}\simeq-c^{2}dt^{2}+dr^{2}%
+r^{2}(d\theta^{2}+\sin^{2}\theta d\phi^{2})$. With a mass $M,$ the observer
is in free fall towards the mass with radial velocity $\vec{v}=-\hat{r}\left(
2GM/r\right)  ^{1/2}$ at $r$. Galilean transformation to a uniformly
accelerated frame from the homogeneous, isotropic metric is $t^{\prime
}=t,~r^{\prime}=r+vt$ with $v=\left(  2GM/r\right)  ^{1/2}$ and we get the
metric in the observer frame from%
\begin{equation}
ds^{\prime2}=-c^{2}dt^{2}(1-2GM/c^{2}r)+dr^{2}+2\left(  2GM/r\right)
^{1/2}drdt+r^{2}(d\theta^{2}+\sin^{2}\theta d\phi^{2})
\end{equation}
Though not evident, this is the Schwarzchild metric in a non-diagonal form,
without any divergence at $r=2GM/c^{2}$. \ To get the metric in the usual
coordinates, the $drdt$ mixed term can be eliminated by introducing a new time
coordinate defined by \cite{Taylor-Wheeler},%
\begin{equation}
dT=dt-dr\frac{\left(  2GM/r\right)  ^{1/2}}{c^{2}(1-2GM/c^{2}r)}%
\end{equation}
Then,
\begin{align}
ds^{\prime2} &  =-c^{2}dT^{2}(1-2GM/c^{2}r)+\frac{dr^{2}\left(  2GM/c^{2}%
	r\right)  }{(1-2GM/c^{2}r)}-2\left(  2GM/r\right)  ^{1/2}drdt\\
&  +2\left(  2GM/r\right)  ^{1/2}drdt+dr^{2}+r^{2}(d\theta^{2}+\sin^{2}\theta
d\phi^{2})\\
&  =-c^{2}dT^{2}(1-2GM/c^{2}r)+\frac{dr^{2}}{(1-2GM/c^{2}r)}+r^{2}(d\theta
^{2}+\sin^{2}\theta d\phi^{2})
\end{align}
We have the Schwarzchild line element in the standard form, obtained from a Galilean velocity transformation.


\begin{thebibliography}{99}
\bibitem {Bhabha}H. J. Bhabha, Nuclear forces, heavy electrons, and the
$\beta$-decay, Nature \textbf{141}, 117 (1938); this paper was written when
muons were misidentified with the Yukawa particles and it was the first time
such a decaying nuclear particle was treated as a relativistic clock.
\end{thebibliography}
\end{document}